\documentclass[11pt,a4paper]{article}
\usepackage{amsmath}
\usepackage{amssymb}
\usepackage{mathtools}
\usepackage{bm}

\usepackage{algorithm}
\usepackage{algpseudocode}

\usepackage{booktabs}
\usepackage{graphicx}
\usepackage{multirow}
\usepackage{tikz}
\usetikzlibrary{arrows.meta, positioning, fit, calc, decorations.pathreplacing}
\usepackage{pgfplots}
\pgfplotsset{compat=1.18}

\usepackage{siunitx}

\usepackage{listings}
\usepackage{xcolor}
\definecolor{quantum-blue}{RGB}{31,119,180}
\definecolor{classical-red}{RGB}{214,39,40}

\usepackage{microtype}   % character protrusion + font expansion; fixes text overfull hboxes

\newcommand{\orcid}[1]{\href{https://orcid.org/#1}{\textcolor{quantum-blue}{ORCID}}}

\newcommand{\hilbert}{\mathcal{H}}

\newcommand{\unitary}{\hat{U}}

\newcommand{\identity}{\mathbb{I}}
\newcommand{\BigO}[1]{\mathcal{O}(#1)}
\newcommand{\ket}[1]{|#1\rangle}

\newcommand{\ketbra}[2]{|#1\rangle\langle#2|}

\DeclareMathOperator*{\argmax}{arg\,max}

\newcommand{\QAOA}{QAOA}
\newcommand{\QUBO}{QUBO}
\newcommand{\POMDP}{POMDP}
\newcommand{\MHT}{MHT}
\newcommand{\MTDA}{MTDA}
\newcommand{\QBRL}{QBRL}
\newcommand{\BIQAE}{BIQAE}
\newcommand{\QANTIS}{QANTIS}
\newcommand{\ZNE}{ZNE}
\newcommand{\PEC}{PEC}
\newcommand{\NISQ}{NISQ}
\newcommand{\FTQC}{FTQC}

\usepackage[margin=1in]{geometry}
\usepackage{authblk}
\usepackage[numbers]{natbib}
\usepackage{subcaption}
\usepackage{hyperref}
\hypersetup{colorlinks=true,linkcolor=blue,citecolor=blue,urlcolor=blue}

\newcommand{\paperabstract}{%
Autonomous navigation under uncertainty requires solving partially
observable Markov decision processes (POMDPs) for planning and
assigning sensor measurements to tracked targets---a task known as
multi-target data association (MTDA).  Both problems become
computationally demanding at scale: belief conditioning costs
$\mathcal{O}(P(e)^{-1})$ per node under rare evidence, while MTDA is
NP-hard.  Quantum amplitude amplification can quadratically reduce
the belief-update query cost to $\mathcal{O}(P(e)^{-1/2})$, while
QUBO reformulations expose MTDA to quantum and quantum-inspired
optimisation heuristics.  We present QANTIS, a modular platform
that integrates quantum belief update (Grover amplitude amplification
and BIQAE), QUBO-based data association via FPC-QAOA, and composable
error mitigation, and we report a 45-experiment hardware study on
three IBM Heron backends.
On hardware, a single Grover iterate applied to a Tiger belief oracle
amplifies a rare observation probability from $0.179$ to $0.907$
($5.1\!\times$; ISA~18) while preserving the Bayesian posterior
(Hellinger~$0.0015$), increasing usable-shot yield from $1{,}463$ to
$7{,}429$.
We interpret this as a hardware validation of the quadratic
query-complexity mechanism at $k{=}1$ with posterior preservation,
rather than a wall-clock advantage claim.
We further demonstrate, to our knowledge, the first closed-loop
hybrid quantum--classical Tiger POMDP on superconducting hardware
($T{=}8$, max Hellinger below $0.015$), and empirically characterise
NISQ feasibility boundaries: ZNE-based error mitigation is beneficial
below ISA~${\approx}100$ and harmful above ISA~${\gtrsim}1{,}000$;
FPC-QAOA is meaningful at ${\leq}15$~QUBO variables
(ISA~${\lesssim}450$).
These results characterise practical operating regimes on current
superconducting hardware rather than wall-clock quantum advantage
at today's problem scales.%
}

\newcommand{\paperkeywords}{quantum computing, POMDP planning, multi-target data association, QAOA, QUBO, amplitude amplification, belief update, NISQ, autonomous navigation}

\title{\textbf{QANTIS: A Hardware-Validated Quantum Platform for POMDP Planning and Multi-Target Data Association}}

\author[1]{Bayram Y\"{u}ksel Eker \orcid{0009-0003-0167-5763}}
\author[2]{Suayb S.~Arslan \orcid{0000-0003-3779-0731}}
\author[1,3]{\"{O}zg\"{u}r Nazl{\i} \orcid{0009-0009-0784-0823}}
\author[4]{Mustafa Serhat Demirgil \orcid{0009-0006-4310-3546}}
\author[5]{Furkan Delig\"{o}z \orcid{0009-0008-7277-8408}}
\affil[1]{Neura Parse Ltd., London, United Kingdom}
\affil[2]{Department of Computer Engineering, Bo\u{g}azi\c{c}i University, Istanbul, Turkey}
\affil[3]{\. Izmir Ekonomi \"{U}niversitesi, \. Izmir, Turkey}
\affil[4]{Faculty of Engineering, University of Toronto, Toronto, Canada}
\affil[5]{Istanbul Technical University, Istanbul, Turkey}
\affil[ ]{Correspondence: \texttt{research@neuraparse.com}}

\date{February 2026}

\begin{document}

\maketitle

\begin{abstract}
\paperabstract
\end{abstract}


% !TEX root = ../main-arxiv.tex
%----------------------------------------------------------------------
\section{Introduction}
\label{sec:introduction}
%----------------------------------------------------------------------

Autonomous systems operating under partial observability face two interlocking computational
bottlenecks: (i)~conditioning belief states on rare-evidence observations in
\POMDP{} planning, and (ii)~solving NP-hard multi-target data association at
each sensor frame.  Classical methods scale as $\BigO{P(e)^{-1}}$ for rare
observations and $\BigO{n^3}$ per frame (Hungarian) escalating to an exponentially
growing hypothesis tree across scans---both become real-time infeasible for
dense environments and deep horizons.

\paragraph{The quantum opportunity.}
Amplitude amplification~\cite{brassard_amplification_2002} reduces the
per-belief-node cost to $\BigO{P(e)^{-1/2}}$---a provable quadratic
speedup achievable in a fault-tolerant setting.  \QAOA{}~\cite{farhi_qaoa_2014}
and fixed-parameter-count variants~\cite{saavedra_fpcqaoa_2025} encode
the \MTDA{} cost matrix as a \QUBO{} and search for low-energy assignments
on near-term superconducting hardware.  While full fault-tolerance remains a
medium-term prospect, current \NISQ{} hardware (IBM Heron~R2, 156 qubits) can
execute the components that constitute future fault-tolerant implementations,
and simultaneously delineate the feasibility boundary for near-term quantum utility.

\paragraph{Contributions.}
We present \QANTIS{} (\emph{Quantum Autonomous Navigation, Tracking \& Intelligence
System}), a modular platform that integrates three recent quantum algorithms for
autonomous systems and, to our knowledge, provides their first joint hardware
validation on IBM Heron processors.
\QANTIS{} makes four contributions:

\begin{enumerate}
  \item \textbf{Hardware demonstration of Grover-AA on a \POMDP{} belief oracle.}
    We validate the \QBRL{} hybrid planner of
    Cunha~et~al.~\cite{cunha_hybrid_2025} on IBM Heron.  A single Grover
    iterate on the Tiger belief oracle boosts $P(\text{rare obs})$ from
    $0.179$ to $0.907$ ($5.1\!\times$; ISA~18) while preserving the Bayesian
    posterior (Hellinger~$0.0015$).

  \item \textbf{Closed-loop hybrid quantum--classical \POMDP{} with \BIQAE{}.}
    We demonstrate, to our knowledge, the first closed-loop hybrid
    quantum--classical Tiger \POMDP{} on superconducting hardware
    ($T{=}8$; max Hellinger~$0.0149$), combining quantum belief update
    and \BIQAE{} amplitude estimation~\cite{li_biqae_2026} with classical
    \QBRL{} action selection.  \BIQAE{} is validated across seven
    amplitudes, two oracle types, and three IBM Heron QPUs (95\% CI
    coverage at all amplitudes).

  \item \textbf{Empirical \NISQ{} feasibility boundary for FPC-\QAOA{} \MTDA{}.}
    We deploy FPC-\QAOA{}~\cite{saavedra_fpcqaoa_2025} on IBM Heron~R2
    across five circuit depths and two instance sizes, establishing an
    empirical boundary at ${\approx}11$--$15$ \QUBO{} variables and
    ISA~$\lesssim\!450$ (variable-count-driven).  These results should be
    interpreted as a feasibility study; Hungarian is optimal and fast at
    all tested scales.

  \item \textbf{Composable error mitigation pipeline and \NISQ{} boundary.}
    \ZNE{}~\cite{temme_error_mitigation_2017} and Pauli twirling + XY4
    dynamical decoupling~\cite{campbell_frame_randomization_2025} are
    benchmarked, yielding a four-point ISA depth boundary (near-noiseless
    $\leq\!20$ / effective $\leq\!100$ / marginal $400$--$1\,000$ / harmful
    $>\!1\,000$) and a warm-start transfer guideline (benefits hardware at
    simulator convergence $\geq\!95\%$).
\end{enumerate}

The platform is implemented in three composable Python packages
(\texttt{quantum-common}, \texttt{quantum-pomdp}, \texttt{quantum-mht})
built on Qiskit~\cite{qiskit_2024} and PennyLane~\cite{bergholm_pennylane_2022}.

\paragraph{Hardware pilot highlights.}
A 45-experiment pilot on three IBM Heron QPUs
(\textsc{ibm\_torino}~R1, \textsc{ibm\_fez}~R2, \textsc{ibm\_marrakesh}~R2)
yields three headline results (details in Section~\ref{ssec:hardware-pilot}):
\textit{(i)}~Grover-AA (ISA~18) amplifies $P(\text{rare obs})$ from
$0.179$ to $\mathbf{0.907}$ ($5.1\!\times$) with posterior preservation
(Hellinger~$0.0015$), validating the query-complexity mechanism at
$k{=}1$ (not a wall-clock advantage claim);
\textit{(ii)}~closed-loop hybrid quantum--classical \POMDP{} at $T{=}4$
and $T{=}8$ (max Hellinger~$0.0149$, PASS), with four independent $T{=}4$
replications all PASS;
\textit{(iii)}~4-state Tiger ($|S|{=}4$, ISA~162, Hellinger~$0.044$, PASS).
To our knowledge, no prior work reports a closed-loop quantum \POMDP{}
execution (Tiger domain) on IBM Heron superconducting
hardware~\cite{aaronson_qpomdp_2014,cunha_hybrid_2025}.

\paragraph{Paper organisation.}
Section~\ref{sec:background} fixes notation.
Section~\ref{sec:related-work} positions \QANTIS{} relative to 2026 literature.
Sections~\ref{sec:problem-formulation}--\ref{sec:quantum-mht} present the algorithms.
Section~\ref{sec:implementation} describes the software architecture.
Section~\ref{sec:experiments} reports experimental results.
Sections~\ref{sec:discussion}--\ref{sec:conclusion} discuss limitations and conclude.

% !TEX root = ../main-arxiv.tex
%----------------------------------------------------------------------
\section{Background and Notation}
\label{sec:background}
%----------------------------------------------------------------------

We briefly fix notation and state the key results that \QANTIS{} builds upon.
Table~\ref{tab:notation} consolidates all symbols used in the paper.

%----------------------------------------------------------------------
\subsection{Quantum Computing Primitives}
\label{ssec:qc}
%----------------------------------------------------------------------

An $n$-qubit register inhabits $\hilbert^{\otimes n} \cong \mathbb{C}^{2^n}$;
a state $\ket{\psi} = \sum_x \alpha_x\ket{x}$ is described by amplitudes
$\alpha_x \in \mathbb{C}$ with $\sum_x |\alpha_x|^2 = 1$.
Measurement in the computational basis yields outcome $x$ with probability
$|\alpha_x|^2$~\cite{nielsen_quantum_2010}.

\paragraph{Amplitude amplification.}
Given a unitary $\mathcal{A}$ that prepares a state with \emph{good}-subspace
amplitude $\sqrt{a}$ ($a = P(e)$), the Grover iterate
$G = \mathcal{A}\,S_0\,\mathcal{A}^\dagger\,S_e$
(where $S_e$ marks good states and $S_0 = \identity - 2\ketbra{0^n}{0^n}$)
boosts the success probability to near unity in
\begin{equation}
  k^\star = \BigO{a^{-1/2}} \text{ oracle calls},
  \label{eq:aa-cost}
\end{equation}
a quadratic speedup over classical rejection sampling at
$\BigO{a^{-1}}$~\cite{brassard_amplification_2002}.
This speedup holds in the fault-tolerant regime; the \NISQ{} boundary is
characterised empirically in Section~\ref{ssec:advantage-thresholds}.

\paragraph{QAOA.}
The $p$-layer \QAOA{} ansatz~\cite{farhi_qaoa_2014} prepares
$\ket{\bm\gamma,\bm\beta} = \prod_{l=1}^p e^{-i\beta_l H_M}e^{-i\gamma_l H_C}\ket{+}^{\otimes n}$,
where $H_C$ encodes the cost function and $H_M = \sum_i X_i$ is the mixer.
Parameters $(\bm\gamma,\bm\beta)$ are optimised classically;
FPC-\QAOA{}~\cite{saavedra_fpcqaoa_2025} fixes the parameter count
independently of $p$ to mitigate barren plateaus.

%----------------------------------------------------------------------
\subsection{\POMDP{} Planning}
\label{ssec:pomdp-background}
%----------------------------------------------------------------------

A \POMDP{} $\mathcal{M} = (S, A, T, \Omega, O, R, \gamma)$ requires the agent
to maintain a \emph{belief state} $b \in \Delta^{|S|-1}$ and update it via
Bayes' rule after action $a$ and observation $o$:
\begin{align}
  \mathcal{B}(b,a,o)(s') &= \frac{O(o|s',a)\sum_s T(s'|s,a)\,b(s)}{P(e)},
  \label{eq:belief-update}\\
  P(e) &= \sum_{s'}O(o|s',a)\sum_s T(s'|s,a)\,b(s).
  \notag
\end{align}
Online planners (\textsc{pomcp}~\cite{silver_pomcp_2010},
\textsc{despot}~\cite{ye_despot_2017}) sample from $b$ to evaluate
$\mathcal{B}$, requiring $\BigO{P(e)^{-1}}$ samples per belief node.
Amplitude amplification (Eq.~\ref{eq:aa-cost}) reduces this to
$\BigO{P(e)^{-1/2}}$, motivating the quantum planner in
Section~\ref{sec:quantum-pomdp}.

%----------------------------------------------------------------------
\subsection{Multi-Hypothesis Tracking}
\label{ssec:mht-background}
%----------------------------------------------------------------------

Given $N$ tracks with Kalman-predicted states $\{(\bm{x}_i^-, P_i^-)\}$
and $M$ measurements $\{\bm{z}_j\}$, the association cost is the negative
log-likelihood $c_{ij} = \tfrac12(d_{ij}^2 + \ln\det(2\pi S_{ij}))$,
where $d_{ij}^2 = \bm\nu_{ij}^\top S_{ij}^{-1}\bm\nu_{ij}$ is the squared
Mahalanobis distance and $S_{ij} = \bm{H}P_i^-\bm{H}^\top + R$ is the
innovation covariance~\cite{barshalom_tracking_1995}.
The optimal single-frame assignment is solved in $\BigO{n^3}$ by the
Hungarian algorithm~\cite{kuhn_hungarian_1955}; the multi-scan extension
(\MHT{}) is NP-hard~\cite{reid_mht_1979}.  Encoding assignment variables
$x_{ij}\in\{0,1\}$ as a \QUBO{} instance allows quantum solvers to target
this bottleneck (Section~\ref{sec:quantum-mht}).

%----------------------------------------------------------------------
\subsection{Notation}
\label{ssec:notation}
%----------------------------------------------------------------------

Throughout, $P(e)$ denotes the evidence probability $P(o \mid b, a)$;
$\mathcal{B}(b,a,o)$ is the Bayesian belief update operator
(Eq.~\ref{eq:belief-update}); ISA denotes the compiled two-qubit-gate
depth (post-transpilation on the target backend) used to characterise
\NISQ{} feasibility; and ``PASS'' indicates that the Hellinger distance
between hardware and exact posteriors is below $0.15$
(${\leq}1.1\%$ total-variation distance), as defined in
Section~\ref{ssec:hardware-pilot}.

\begin{table}[t]
  \centering
  \caption{Principal notation.}
  \label{tab:notation}
  \resizebox{\columnwidth}{!}{%
  \begin{tabular}{@{}cl|cl@{}}
    \toprule
    \textbf{Symbol} & \textbf{Description} &
    \textbf{Symbol} & \textbf{Description} \\
    \midrule
    $S,A,\Omega$      & State/action/observation spaces &
    $P(e)$            & Evidence probability $P(o|b,a)$ \\
    $T,O,R,\gamma$    & Transition/obs/reward/discount &
    $\mathcal{B}$     & Bayesian belief update operator \\
    $b\in\Delta^{|S|-1}$ & Belief state &
    $H$               & Planning horizon \\
    $N,M$             & Track/measurement count &
    $x_{ij}\in\{0,1\}$ & Assignment variable \\
    $\bm{x}_i^-,P_i^-$ & Predicted state and covariance &
    $\lambda$         & \QUBO{} penalty coefficient \\
    $d_{ij}^2$        & Mahalanobis distance &
    $c_{ij}$          & Association cost (NLL) \\
    $k^\star$         & Optimal Grover iterations &
    $p$               & \QAOA{} depth \\
    $\bm\gamma,\bm\beta$ & \QAOA{} parameters &
    $H_C,H_M$         & Cost/mixer Hamiltonian \\
    \bottomrule
  \end{tabular}}% end resizebox
\end{table}

% !TEX root = ../main-arxiv.tex
%----------------------------------------------------------------------
\section{Related Work}
\label{sec:related-work}
%----------------------------------------------------------------------

%----------------------------------------------------------------------
\subsection{Quantum Approaches to Planning Under Uncertainty}
\label{ssec:quantum-decision}
%----------------------------------------------------------------------

\textbf{Quantum \POMDP{} theory.}
Aaronson~\cite{aaronson_qpomdp_2014} established complexity-theoretic
separations for quantum \POMDP{} planning but did not prescribe \NISQ{}-era
algorithms.  Cunha~et~al.~\cite{cunha_hybrid_2025} proposed \QBRL{},
the first practical hybrid planner using amplitude amplification for the
root belief update at $\BigO{P(e)^{-1/2}}$ cost.  Li~et~al.\ published
the \BIQAE{} adaptive amplitude estimation protocol in
\emph{Quantum}~\cite{li_biqae_2026}, showing Heisenberg-scaling
credible intervals; Ramoa~et~al.~\cite{ramoa_bqae_2025} analysed
exponentially growing $K$-schedules as near-optimal.
\QANTIS{} integrates \QBRL{} and \BIQAE{} and provides---to our
knowledge---the first IBM Heron hardware demonstration of a quantum
\POMDP{} loop on superconducting hardware (Section~\ref{par:e2e-pomdp}).

\textbf{Classical \POMDP{} solvers.}
\textsc{pomcp}~\cite{silver_pomcp_2010} and
\textsc{despot}~\cite{ye_despot_2017} achieve state-of-the-art planning
quality but both scale as $\BigO{P(e)^{-1}}$ per belief update node,
a bottleneck our quantum approach targets.

%----------------------------------------------------------------------
\subsection{Quantum Optimisation for Tracking}
\label{ssec:quantum-tracking}
%----------------------------------------------------------------------

\textbf{\QUBO{}-based data association.}
Stollenwerk~et~al.~\cite{stollenwerk_atm_2021} first cast \MTDA{} as a
\QUBO{} for D-Wave annealing.  Ihara~\cite{ihara_quantum_tracking_2025}
demonstrated warm-start reverse annealing for temporally correlated frames.
FPC-\QAOA{}~\cite{saavedra_fpcqaoa_2025} extends this line to gate-model
hardware with fixed variational parameter count.
\QANTIS{} adopts the \QUBO{} encoding of Stollenwerk~et~al.\ and
extends it to FPC-\QAOA{} on IBM Heron hardware.

\textbf{HHL-based quantum tracking (2026).}
Chiotopoulos~et~al.~\cite{chiotopoulos_trackhhl_2026} propose TrackHHL,
an HHL-variant with $\BigO{\sqrt{N}\log N}$ gate complexity for particle
trajectory reconstruction, benchmarked on IBM Heron and Quantinuum~H2
noise models.  TrackHHL targets a different regime (linear-system
solvers for Kalman filters) and uses no \QUBO{}/\QAOA{} encoding; our \QUBO{}
formulation handles the combinatorial assignment problem directly.

%----------------------------------------------------------------------
\subsection{\QAOA{} and Error Mitigation on \NISQ{} Hardware (2026)}
\label{ssec:qa-nisq-2026}
%----------------------------------------------------------------------

\textbf{\QAOA{} training efficiency.}
Jang~et~al.~\cite{jang_orbit_qaoa_2026} (Orbit-\QAOA{}, 2026) achieve
an $81.8\%$ reduction in training steps via cyclic layerwise parameter
freezing.  Nzongani~et~al.~\cite{torrontegui_schedule_transfer_2026}
(2026) reduce the \QAOA{} parameter space from $2p$ to two global
hyperparameters by adiabatic schedule transfer, matching our FPC approach's
motivation.

\textbf{\QAOA{}+ZNE on IBM hardware.}
Ribeiro~\cite{ribeiro_qaoa_zne_2026} validates \QAOA{}+ZNE on
\textsc{ibm\_fez} and \textsc{ibm\_torino} (same processors as ours)
for an 88-variable portfolio problem, achieving a $31.6\%$ improvement
over a classical greedy baseline ($p_\mathrm{stat}{=}0.0009$, Cohen's $d{=}2.01$).
Our 11-variable FPC-\QAOA{} achieves $64.1\%\pm3.3\%$ of the optimal
(Hungarian) baseline---a more demanding metric since Hungarian is
optimal, not heuristic.  Together, these results confirm the practical
\NISQ{} feasibility regime: both works find meaningful quantum signal below
ISA~$\sim\!450$ and noise domination above.

\textbf{IBM Heron hardware benchmarking.}
Kiiamov and Tayurskii~\cite{kiiamov_heron2_2026} benchmark IBM Heron~2
for quantum simulation, confirming $<7\%$ relative error on 6-qubit circuits.
Marshall~et~al.~\cite{marshall_qmcmc_2026} demonstrate quantum-enhanced
MCMC on 117 variables on IBM hardware, establishing the current
state-of-the-art for IBM combinatorial optimisation.
Our work extends to quantum \POMDP{} (an entirely different algorithmic
class) and establishes the first multi-step closed-loop hardware
demonstration combining belief update with action selection.

%----------------------------------------------------------------------
\subsection{Gap Analysis}
\label{ssec:gap}
%----------------------------------------------------------------------

\begin{table}[t]
  \centering
  \caption{Positioning of \QANTIS{} relative to prior and concurrent
    work.  Q-Plan: quantum \POMDP{} planning.  Q-DA: quantum data association.
    HW: real quantum hardware experiment.  EM: error mitigation.
    CL: closed-loop action selection.}
  \label{tab:gap-analysis}
  \resizebox{\columnwidth}{!}{%
  \begin{tabular}{@{}lccccc@{}}
    \toprule
    \textbf{System} & \textbf{Q-Plan} & \textbf{Q-DA}
      & \textbf{HW} & \textbf{EM} & \textbf{CL} \\
    \midrule
    Aaronson (2014)~\cite{aaronson_qpomdp_2014}
      & \checkmark & -- & -- & -- & -- \\
    Stollenwerk et~al.\ (2021)~\cite{stollenwerk_atm_2021}
      & -- & \checkmark & \checkmark & -- & -- \\
    Cunha et~al.\ (2025)~\cite{cunha_hybrid_2025}
      & \checkmark & -- & -- & -- & -- \\
    Ribeiro (2026)~\cite{ribeiro_qaoa_zne_2026}
      & -- & \checkmark & \checkmark & \checkmark & -- \\
    TrackHHL (2026)~\cite{chiotopoulos_trackhhl_2026}
      & -- & \checkmark & -- & -- & -- \\
    Marshall et~al.\ (2026)~\cite{marshall_qmcmc_2026}
      & -- & \checkmark & \checkmark & -- & -- \\
    \midrule
    \textbf{\QANTIS{} (this work)}
      & \checkmark & \checkmark & \checkmark & \checkmark & \checkmark \\
    \bottomrule
  \end{tabular}}% end resizebox
\end{table}

Table~\ref{tab:gap-analysis} summarises the landscape.  We are not
aware of a single system that jointly provides quantum \POMDP{} planning,
quantum data association, hardware validation, error mitigation, and
closed-loop action selection.  \QANTIS{} addresses this gap
with---to our knowledge---the first hardware-validated
quantum--classical perception--decision pipeline on IBM Heron processors.

% !TEX root = ../main-arxiv.tex
%----------------------------------------------------------------------
\section{Problem Formulation}
\label{sec:problem-formulation}
%----------------------------------------------------------------------

%----------------------------------------------------------------------
\subsection{\POMDP{} Planning}
\label{ssec:pomdp-problem}
%----------------------------------------------------------------------

Given $\mathcal{M} = (S, A, T, \Omega, O, R, \gamma)$ and initial belief
$b_0$, the planning problem is to compute the optimal value function
\begin{equation}
  V^\star(b) = \max_{a\in A}\Bigl[R(b,a) + \gamma\sum_{o\in\Omega}P(o|b,a)\,V^\star(\mathcal{B}(b,a,o))\Bigr],
  \label{eq:bellman}
\end{equation}
where the belief update $\mathcal{B}$ and evidence probability $P(e)$ are as
in Eq.~\eqref{eq:belief-update}.  Online planners evaluate
Eq.~\eqref{eq:bellman} via Monte Carlo sampling from $b$, requiring
$C_\text{cl} = P(e)^{-1}$ trials per accepted sample.  Over a tree of depth
$H$ with branching factor $|A||\Omega|$, the total cost scales as
$\BigO{(|A||\Omega|)^H P(e)^{-1}}$.  Amplitude amplification reduces the
per-node cost to $P(e)^{-1/2}$ (Eq.~\ref{eq:aa-cost}), targeting a quadratic
improvement in the evidence-probability term.

%----------------------------------------------------------------------
\subsection{Multi-Target Data Association (\MTDA{})}
\label{ssec:mtda-problem}
%----------------------------------------------------------------------

Given $N$ tracks and $M$ measurements with association costs $c_{ij}$ (see
Section~\ref{ssec:mht-background}), the \MTDA{} problem is
\begin{equation}
  \min_{\{x_{ij}\in\{0,1\}\}} \sum_{i,j} c_{ij}\,x_{ij}
  \;\;\text{s.t.}\;\;
  \sum_j x_{ij}\leq 1\;\forall i,\quad
  \sum_i x_{ij}\leq 1\;\forall j.
  \label{eq:mtda}
\end{equation}
Converting to an unconstrained \QUBO{} via quadratic penalties:
\begin{equation}
  H_\QUBO{} = \sum_{i,j}c_{ij}x_{ij}
    + \lambda\sum_i\!\Bigl(\sum_j x_{ij}-1\Bigr)^{\!2}
    + \lambda\sum_j\!\Bigl(\sum_i x_{ij}-1\Bigr)^{\!2},
  \label{eq:qubo}
\end{equation}
with penalty coefficient $\lambda = 1.5\max_{ij}|c_{ij}|$ (following
Stollenwerk~et~al.~\cite{stollenwerk_atm_2021}).  The Ising mapping
$x_{ij}=(1-z_{ij})/2$ converts Eq.~\eqref{eq:qubo} to the problem
Hamiltonian $H_C$ for \QAOA{}.

%----------------------------------------------------------------------
\subsection{Unified Autonomous System Model}
\label{ssec:unified-model}
%----------------------------------------------------------------------

\QANTIS{} closes the perception--decision loop in four stages at each
time step~$t$: (1)~raw measurements $\{\bm{z}_j^{(t)}\}$ are ingested
by the \MHT{} module; (2)~FPC-\QAOA{} solves~\eqref{eq:qubo} to
associate measurements with tracks and update state estimates
$\{\hat{\bm{x}}_i^{(t)}\}$; (3)~track estimates are discretised to
form belief~$b_t$; and (4)~\QBRL{} queries the quantum belief update
circuit to select action $a_t^\star$ maximising the estimated
$Q^\star(b_t,\cdot)$.  Either component can run classically when hardware
is unavailable or problem size exceeds the \NISQ{} feasibility boundary
(default fallback: $n_\text{var}{>}15$ or predicted ISA${>}450$
triggers the classical solver automatically; predicted ISA is obtained
from a transpilation preview pass prior to job submission).

% !TEX root = ../main-arxiv.tex
%----------------------------------------------------------------------
\section{Quantum-Enhanced \POMDP{} Planning}
\label{sec:quantum-pomdp}
%----------------------------------------------------------------------

%----------------------------------------------------------------------
\subsection{Quantum Belief Update Circuit}
\label{ssec:belief-circuit}
%----------------------------------------------------------------------

The belief update circuit operates on five quantum registers encoding the
\POMDP{} dimensions:

\begin{center}
\begin{tabular}{@{}clc@{}}
  \toprule
  \textbf{Register} & \textbf{Semantics} & \textbf{Qubits} \\
  \midrule
  $S_t$     & Current state   & $\lceil\log_2|S|\rceil$ \\
  $A_t$     & Action          & $\lceil\log_2|A|\rceil$ \\
  $S_{t+1}$ & Next state      & $\lceil\log_2|S|\rceil$ \\
  $O_{t+1}$ & Observation     & $\lceil\log_2|\Omega|\rceil$ \\
  $R_{t+1}$ & Reward (ancilla)& $k_5$ precision bits \\
  \bottomrule
\end{tabular}
\end{center}

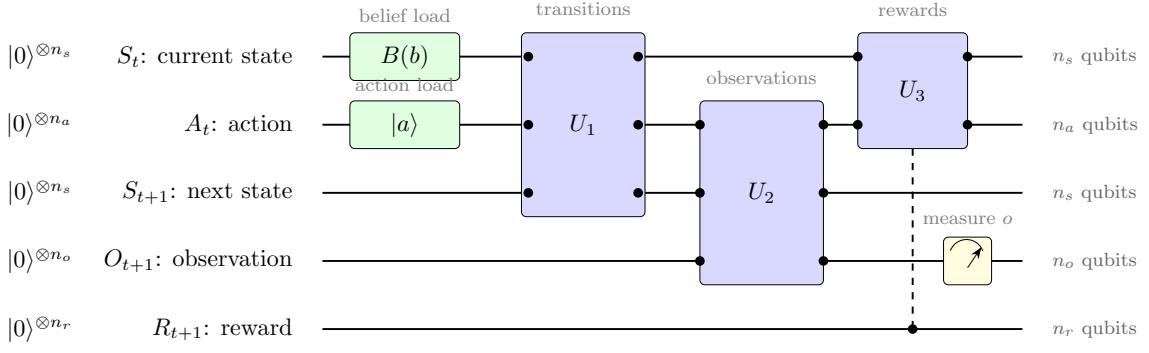
\begin{figure*}[t]
\centering
\resizebox{0.95\textwidth}{!}{%
\begin{tikzpicture}[
  reg/.style={font=\small},
  gate/.style={draw, fill=blue!10, minimum width=1.6cm, minimum height=0.7cm,
    rounded corners=2pt, font=\small},
  meas/.style={draw, fill=yellow!15, minimum width=0.7cm, minimum height=0.7cm,
    rounded corners=2pt, font=\small},
  wire/.style={thick},
  >=Stealth,
]
  % Register labels (left side)
  \def\vgap{1.0}
  \foreach \i/\lbl/\nbits in {
    0/{$S_t$: current state}/n_s,
    1/{$A_t$: action}/n_a,
    2/{$S_{t+1}$: next state}/n_s,
    3/{$O_{t+1}$: observation}/n_o,
    4/{$R_{t+1}$: reward}/n_r} {
    \pgfmathsetmacro{\y}{-\i*\vgap}
    \node[reg, anchor=east] at (-0.3, \y) {\lbl};
    \node[font=\scriptsize, text=black!60, anchor=west] at (10.5, \y) {$\nbits$ qubits};
    % Wires
    \draw[wire] (0, \y) -- (10.2, \y);
    % Initial state
    \node[font=\small, anchor=east] at (-3.5, \y) {$\ket{0}^{\otimes \nbits}$};
  }

  % Belief loading block
  \node[gate, fill=green!12, minimum height=0.7cm] at (1.2, 0) {$B(b)$};
  \node[font=\scriptsize, text=black!50] at (1.2, 0.6) {belief load};

  % Action loading block
  \node[gate, fill=green!12, minimum height=0.7cm] at (1.2, -\vgap) {$\ket{a}$};
  \node[font=\scriptsize, text=black!50] at (1.2, -\vgap+0.6) {action load};

  % U1: Transition encoding
  \node[gate, fill=blue!15, minimum height=2.7cm, minimum width=1.8cm]
    (u1) at (3.8, -\vgap) {$U_1$};
  \node[font=\scriptsize, text=black!50] at (3.8, -\vgap+1.7) {transitions};
  % Connection dots for U1
  \foreach \y in {0, -\vgap, -2*\vgap} {
    \fill (3.0, \y) circle (2pt);
    \fill (4.6, \y) circle (2pt);
  }

  % U2: Observation encoding
  \node[gate, fill=blue!15, minimum height=2.7cm, minimum width=1.8cm]
    (u2) at (6.4, -2*\vgap) {$U_2$};
  \node[font=\scriptsize, text=black!50] at (6.4, -2*\vgap+1.7) {observations};
  \foreach \y in {-\vgap, -2*\vgap, -3*\vgap} {
    \fill (5.5, \y) circle (2pt);
    \fill (7.3, \y) circle (2pt);
  }

  % U3: Reward encoding
  \node[gate, fill=blue!15, minimum height=1.7cm, minimum width=1.6cm]
    (u3) at (8.6, -0.5*\vgap) {$U_3$};
  \node[font=\scriptsize, text=black!50] at (8.6, -0.5*\vgap+1.2) {rewards};
  \foreach \y in {0, -\vgap} {
    \fill (7.8, \y) circle (2pt);
    \fill (9.4, \y) circle (2pt);
  }
  % U3 also connects to reward register
  \draw[wire, dashed] (8.6, -0.5*\vgap-0.85) -- (8.6, -4*\vgap);
  \fill (8.6, -4*\vgap) circle (2pt);

  % Measurement on observation register
  \node[meas] (meas) at (9.4, -3*\vgap) {};
  \draw (9.15, -3*\vgap+0.15) arc (160:20:0.25cm);
  \draw[->, thin] (9.4, -3*\vgap-0.1) -- (9.6, -3*\vgap+0.2);
  \node[font=\scriptsize, text=black!50] at (9.4, -3*\vgap+0.6) {measure $o$};

\end{tikzpicture}%
}
\caption{Quantum belief update circuit for a single \POMDP{} step
  (schematic; register sizes $n_s{=}\lceil\log_2|S|\rceil$,
  $n_a{=}\lceil\log_2|A|\rceil$, $n_o{=}\lceil\log_2|\Omega|\rceil$,
  $n_r{=}$ reward precision bits).
  Unitaries $U_1$, $U_2$, $U_3$ encode the transition, observation,
  and reward models respectively.  Measuring the observation register
  and conditioning on outcome~$o$ (via amplitude amplification) yields
  the updated belief in the $S_{t+1}$ register.
  See Appendix~\ref{app:circuit-details} for the full hardware qubit
  allocation.}
\label{fig:belief-circuit}
\end{figure*}

Three unitaries are applied in sequence:
$U_1\ket{s}\ket{a}\ket{0} = \ket{s}\ket{a}\sum_{s'}\!\sqrt{T(s'|s,a)}\ket{s'}$
(transition encoding via uniformly controlled $R_y$ gates);
$U_2\ket{s'}\ket{a}\ket{0} = \ket{s'}\ket{a}\sum_o\!\sqrt{O(o|s',a)}\ket{o}$
(observation encoding);
$U_3$ (reward as phase amplitude).
Given belief $b$ loaded into $S_t$, the full circuit prepares
\begin{equation}
  \ket{\Phi(b,a)} = \sum_{s,s',o}\sqrt{b(s)T(s'|s,a)O(o|s',a)}\;\ket{s}\ket{a}\ket{s'}\ket{o}\ket{r}.
  \label{eq:joint-state}
\end{equation}
Measuring register $O_{t+1}$ and obtaining $o$ collapses $S_{t+1}$ to the
amplitude-encoded posterior $b'=\mathcal{B}(b,a,o)$ with success probability $P(e)$.
Figure~\ref{fig:belief-circuit} depicts the circuit schematically.

% Quantum belief update circuit diagram

%----------------------------------------------------------------------
\subsection{Amplitude Amplification and \BIQAE{}}
\label{ssec:biqae}
%----------------------------------------------------------------------

\paragraph{Amplification.}
Defining the evidence oracle $S_e$ (which phase-flips states with $O_{t+1}{=}o$)
and the Grover iterate $G(o) = \mathcal{A}\,S_0\,\mathcal{A}^\dagger\,S_e$,
applying $G(o)$ a total of $k^\star = \lfloor\pi/(4\arcsin\sqrt{P(e)})-\tfrac12\rfloor$
times boosts the success probability to $\geq 1-P(e)$ at cost
$\BigO{P(e)^{-1/2}}$ circuit evaluations (Eq.~\ref{eq:aa-cost}).

\paragraph{Adaptive \BIQAE{}.}
When $P(e)$ is unknown, \BIQAE{}~\cite{li_biqae_2026} maintains a Bayesian
posterior $\pi_t(\theta)$ over the amplitude angle $\theta=\arcsin\sqrt{P(e)}$.
At each round $t$, it applies $k_t = 3^t$ Grover iterations (exponential
schedule, near-optimal per~\cite{ramoa_bqae_2025}), measures a binary outcome
$m_t$, and updates via Bayes' rule:
$\pi_{t+1}(\theta) \propto \sin^2((2k_t+1)\theta)^{m_t}
  (1-\sin^2((2k_t+1)\theta))^{1-m_t}\,\pi_t(\theta)$.
Termination is declared when the 95\% HPD credible interval width
drops below precision $\epsilon$.  Li~et~al.~\cite{li_biqae_2026} show
\BIQAE{} achieves Heisenberg-scaling ($\BigO{1/N_\text{oracle}}$) and
uses $\approx 14\%$ fewer oracle queries than maximum-likelihood QAE.

Hardware validation on \textsc{ibm\_torino} (1-qubit $R_y(2\theta)\ket{0}$
oracle, ISA~5, 300 shots/iteration) and \textsc{ibm\_marrakesh}
(1-qubit and 2-qubit oracles, ISA 5--11) confirmed all 95\% credible intervals
cover ground truth across seven amplitudes ($a\in[0.10,0.70]$) and all three
backends (Section~\ref{ssec:hardware-pilot}).

%----------------------------------------------------------------------
\subsection{\QBRL{} Hybrid Lookahead Tree}
\label{ssec:qbrl}
%----------------------------------------------------------------------

\QBRL{}~\cite{cunha_hybrid_2025} integrates the quantum belief update at
the \emph{root level only} of a horizon-$H$ search tree; deeper levels use
classical Monte Carlo rollouts.  At the root, for each $(a,o)$ pair:
(i)~\BIQAE{} estimates $P(o|b_0,a)$; (ii)~$k^\star(o)$ Grover iterates
amplify the target observation; (iii)~measurement of $S_{t+1}$ extracts the
classical posterior $b'=\mathcal{B}(b_0,a,o)$; (iv)~classical rollouts
estimate $\hat{V}(b')$.  The action is selected as
$a^\star = \argmax_a [R(b_0,a)+\gamma\sum_o P(o|b_0,a)\hat{V}(\mathcal{B}(b_0,a,o))]$.

\paragraph{Hardware implementation: two complementary \NISQ{} circuits.}
The full 11-qubit belief circuit (ISA~4\,237 after Heron~R2 transpilation)
exceeds the \NISQ{} coherence budget.  We implement two complementary circuits:

\textit{(a)~Grover-AA circuit} (Section~\ref{par:grover-aa}): directly
implements one Grover iterate $G{=}\mathcal{A}\,S_0\,\mathcal{A}^\dagger\,S_e$
on the Tiger belief oracle with ISA depth~18.  Prior $b{=}[0.97,0.03]$,
target obs$=1$ (rare): hardware boosts $P(\text{obs}{=}1)$ from $0.179$
to $0.907$ ($5.1\!\times$).  Post-selecting on obs$=1$ yields hardware
posterior $[0.849,0.151]$ vs.\ classical $[0.851,0.149]$ (Hellinger~$0.0015$),
confirming that Grover amplification \emph{preserves} the correct Bayesian
posterior while achieving $5.1\!\times$ higher usable-shot yield.
This validates the $\BigO{P(e)^{-1/2}}$ query-complexity mechanism on
hardware; it is not a wall-clock advantage claim at current problem scales.

\textit{(b)~Tiger minimal direct-encoding circuit} (ISA~12, ISA~4 for open actions):
the prior is loaded via controlled $R_y$ gates and the observation model applied
with a single conditional $R_y$; post-selection yields the exact Bayesian
posterior.  This circuit achieves $353\times$ depth reduction and is used for
the sequential closed-loop \POMDP{} loops ($T{=}4$ and $T{=}8$)
because it minimises circuit overhead per time step.
Unlike the Grover-AA circuit, it does not amplify the observation probability
(post-selection yield $\approx P(e)$); it is the \emph{architecturally appropriate
\NISQ{} implementation} for sequential inference where $P(e)$ is not rare.
For rare-evidence regimes, the Grover-AA circuit provides the speedup;
integration of Grover-AA into the sequential loop
(one Grover iterate per time step: ISA~18, ${\sim}8\,192$ shots,
${\approx}2$~s wall-clock per step on Heron~R2; feasible within current
hardware budgets) is a direct extension planned for future work.

Section~\ref{par:tiger-minimal} provides further architectural justification.

\paragraph{Complexity summary.}
Realising the full $\BigO{P(e)^{-1/2}}$ speedup for large $k^\star$
requires fault-tolerant gate fidelities ($\epsilon \lesssim 10^{-6}$);
our \NISQ{} demonstration validates the mechanism at $k{=}1$
(Table~\ref{tab:complexity-comparison}).
\begin{table}[h]
  \centering
  \caption{Per-belief-node query cost (theoretical, fault-tolerant regime;
    circuit queries $\propto P(e)^{-1/2}$ require gate fidelities $\lesssim 10^{-6}$
    for large $k^\star$).  Speedup is \QBRL{} over \textsc{pomcp}.
    Hardware demonstration at $P(e){\approx}0.18$: $5.1\!\times$ amplification
    at ISA~18 on \textsc{ibm\_marrakesh} (Section~\ref{par:grover-aa}).}
  \label{tab:complexity-comparison}
  \begin{tabular}{@{}cccr@{}}
    \toprule
    $P(e)$ & \textsc{pomcp} & \QBRL{} root & Speedup \\
    \midrule
    $10^{-2}$ & $100$     & $\approx 10$   & $10\times$ \\
    $10^{-3}$ & $1{,}000$ & $\approx 32$   & $31.6\times$ \\
    $10^{-4}$ & $10{,}000$& $\approx 100$  & $100\times$ \\
    \bottomrule
  \end{tabular}
\end{table}

% !TEX root = ../main-arxiv.tex
%----------------------------------------------------------------------
\section{Quantum Multi-Hypothesis Tracking}
\label{sec:quantum-mht}
%----------------------------------------------------------------------

%----------------------------------------------------------------------
\subsection{\MTDA{} Cost Matrix and \QUBO{} Construction}
\label{ssec:qubo-construction}
%----------------------------------------------------------------------

\noindent\fbox{\parbox{0.96\columnwidth}{%
\textbf{What this section contributes.}
(1)~A \QUBO{} formulation of the multi-target data association problem
    with automatic penalty calibration (Section~\ref{ssec:qubo-construction}).
(2)~An FPC-\QAOA{} ansatz that decouples circuit depth from parameter count
    (Section~\ref{ssec:fpc-qaoa}).
(3)~Qubit-resource estimates identifying the \NISQ{} feasibility boundary
    at ${\approx}11$--$15$ variables / ISA~${\lesssim}450$
    (Section~\ref{ssec:qubit-resources}).
(4)~Our \MTDA{} results constitute an empirical ISA/variable feasibility
    study on \NISQ{} hardware rather than a performance replacement for
    the Hungarian algorithm, which is optimal and fast at all tested
    scales ($N{\leq}3$).}}

Multi-scan data association across $T$ frames is
NP-hard~\cite{blackman_mht_2004}, scaling as $\BigO{e^{N}}$ in the
worst case and becoming classically infeasible for dense, long-horizon
tracking ($N > 40$--$50$, $T > 3$).  FPC-\QAOA{} and quantum
annealing target precisely this regime.
Even in the single-frame case, Hungarian
assignment~\cite{kuhn_hungarian_1955} runs in $\BigO{N^3}$: for
$N\!=\!100$ targets at a 30~Hz UAV frame rate ($33$~ms budget),
Hungarian requires ${\sim}730$~ms---more than $20\!\times$ over-budget
(single-threaded Python, Intel i7-12700K; see
Table~\ref{tab:hungarian-scaling}).  This extrapolation is illustrative;
optimised C++/GPU implementations are faster, but multi-scan NP-hardness
remains the primary computational driver.

For $N$ tracks and $M$ measurements, we construct association costs
$c_{ij} = \tfrac12(d_{ij}^2 + \ln\det(2\pi S_{ij}))$ with Mahalanobis
gating at the $\chi^2_{2,0.99}=9.21$ threshold.  False-alarm and
missed-detection sentinel variables $f_j, m_i \in \{0,1\}$ augment the
binary assignment variables $x_{ij}$, giving
$n_\text{var} = NM + N + M$ total \QUBO{} variables.  The \QUBO{}
Hamiltonian is
\begin{equation}
  H_\QUBO{} = \sum_{i,j}c_{ij}x_{ij} + c_\text{miss}\!\sum_i m_i
    + c_\text{fa}\!\sum_j f_j
    + \lambda(H_\text{row} + H_\text{col}),
  \label{eq:qubo-full}
\end{equation}
where $H_\text{row} = \sum_i(\sum_j x_{ij}+m_i-1)^2$,
$H_\text{col} = \sum_j(\sum_i x_{ij}+f_j-1)^2$, and
$\lambda = 1.5\max_{ij}|c_{ij}|$ (auto-calibration following
Stollenwerk~et~al.~\cite{stollenwerk_atm_2021}).
The Ising mapping $x_{ij}=(1-z_{ij})/2$ converts $H_\QUBO{}$ to the
\QAOA{} cost Hamiltonian $H_C$.

%----------------------------------------------------------------------
\subsection{FPC-\QAOA{} for Data Association}
\label{ssec:fpc-qaoa}
%----------------------------------------------------------------------

Standard \QAOA{} at depth $p$ has $2p$ variational parameters
$(\bm\gamma,\bm\beta)$ whose optimisation landscape suffers from barren
plateaus at large $p$.  FPC-\QAOA{}~\cite{saavedra_fpcqaoa_2025}
parameterises angles as smooth functions of the layer index:
$\gamma_l = \gamma(l/p;\,\bm{a})$, $\beta_l = \beta(l/p;\,\bm{b})$ using
a fixed-degree polynomial or trigonometric schedule controlled by
$k \ll p$ coefficients.  This decouples circuit depth from parameter
count, allowing $p$ to increase without expanding the classical
optimisation landscape.

The warm-start schedule $\gamma(t) = \pi t$, $\beta(t) = \pi/4$ (linear
ramp, constant mixer~\cite{saavedra_fpcqaoa_2025}) provides a high-quality
COBYLA initialisation that mimics the adiabatic limit.
Sim-to-hardware parameter transfer~\cite{patel_qaoa_param_transfer_2026}
(COBYLA-optimised parameters applied directly to hardware) achieves
$71.2\%$ of optimal at $p=2$ in our experiments (Section~\ref{ssec:hardware-pilot})
when simulator convergence $\geq 95\%$; Orbit-\QAOA{}~\cite{jang_orbit_qaoa_2026}
and adiabatic schedule transfer~\cite{torrontegui_schedule_transfer_2026}
provide additional 2026 routes to reduce training overhead further.

%----------------------------------------------------------------------
\subsection{Qubit Resource Analysis}
\label{ssec:qubit-resources}
%----------------------------------------------------------------------

\begin{table}[h]
\centering
\caption{Estimated qubit resources for \MTDA{} \QUBO{} instances.
  Gate-model ISA depths use $p=3$ FPC-\QAOA{} on Heron~R2 heavy-hex routing.
  Our hardware experiments (Section~\ref{ssec:hardware-pilot}) validate
  the $N=2,M=3$ ($n=11$) instance.}
\label{tab:qubit-resources}
\begin{tabular}{@{}rrrrl@{}}
\toprule
$N$ & $M$ & $n_\text{var}$ & ISA ($p=3$) & \NISQ{} feasibility \\
\midrule
2  & 3  & 11  & $\sim$435  & \textbf{Validated} \\
3  & 4  & 19  & $\sim$640  & Marginal ($20.4\%$ at $p=1$) \\
5  & 8  & 53  & $>$2\,000  & Beyond current \NISQ{} \\
10 & 15 & 175 & $\gg$10\,000 & FTQC target \\
\bottomrule
\end{tabular}
\end{table}

The empirical \NISQ{} feasibility boundary lies between $n_\text{var}=11$
and $n_\text{var}=19$ (ISA depth 450--640), consistent with the
variable-count-driven limit established in
Section~\ref{ssec:advantage-thresholds}.  Chi-squared gating typically
eliminates 30--60\% of track--measurement pairs, reducing the effective
\QUBO{} size and extending feasibility to slightly larger instances.

% !TEX root = ../main-arxiv.tex
%----------------------------------------------------------------------
\section{QANTIS Platform Architecture}
\label{sec:implementation}
%----------------------------------------------------------------------

\begin{figure}[t]
\centering
\resizebox{\columnwidth}{!}{%
\begin{tikzpicture}[
  pkg/.style={draw, rounded corners=4pt, minimum width=5.8cm,
    minimum height=2.2cm, align=center, font=\small},
  dep/.style={-{Stealth[length=3mm]}, thick, dashed},
  lbl/.style={font=\footnotesize\ttfamily, text=black!70},
]
  % quantum-common (bottom)
  \node[pkg, fill=blue!8] (common) {
    \textbf{\texttt{quantum-common}}\\[3pt]
    Backend abstraction\\
    Error mitigation pipeline\\
    Configuration \& seed management
  };

  % quantum-pomdp (top left)
  \node[pkg, fill=green!8, above left=1.8cm and -0.3cm of common] (pomdp) {
    \textbf{\texttt{quantum-pomdp}}\\[3pt]
    Belief update circuits\\
    BIQAE \& amplitude amplification\\
    QBRL hybrid planner
  };

  % quantum-mht (top right)
  \node[pkg, fill=orange!8, above right=1.8cm and -0.3cm of common] (mht) {
    \textbf{\texttt{quantum-mht}}\\[3pt]
    QUBO formulation\\
    FPC-QAOA \& QUBO solving\\
    Multi-target tracking pipeline
  };

  % Dependency arrows
  \draw[dep] (pomdp.south) -- node[left, lbl, pos=0.4] {depends on} (common.north west);
  \draw[dep] (mht.south)   -- node[right, lbl, pos=0.4] {depends on} (common.north east);

  % No cross-dependency indicator
  \draw[thick, red!60, {Stealth[length=2.5mm]}-{Stealth[length=2.5mm]}]
    ([yshift=0.3cm]pomdp.east) -- node[above, font=\footnotesize\color{red!60}] {no dependency} ([yshift=0.3cm]mht.west);
  \draw[thick, red!60] ([yshift=0.15cm]pomdp.east) ++(1.2,0) -- ++(0,0.3);

  % Hardware backends below
  \node[below=1.2cm of common, font=\small, align=center] (hw) {
    \begin{tabular}{ccc}
    \tikz\fill[blue!40] (0,0) rectangle (0.25,0.25); IBM Qiskit &
    \tikz\fill[green!40] (0,0) rectangle (0.25,0.25); PennyLane &
    \tikz\fill[purple!40] (0,0) rectangle (0.25,0.25); Qiskit Aer (sim.)
    \end{tabular}
  };
  \draw[dep] (common.south) -- node[right, lbl] {dispatches to} (hw.north);

\end{tikzpicture}%
}
\caption{Modular architecture of the \QANTIS{} platform.  The three
  packages maintain a strict acyclic dependency hierarchy:
  \texttt{quantum-pomdp} and \texttt{quantum-mht} depend on
  \texttt{quantum-common} but have no cross-dependency.  The backend
  abstraction layer in \texttt{quantum-common} dispatches to three
  quantum hardware/simulator backends.}
\label{fig:architecture}
\end{figure}
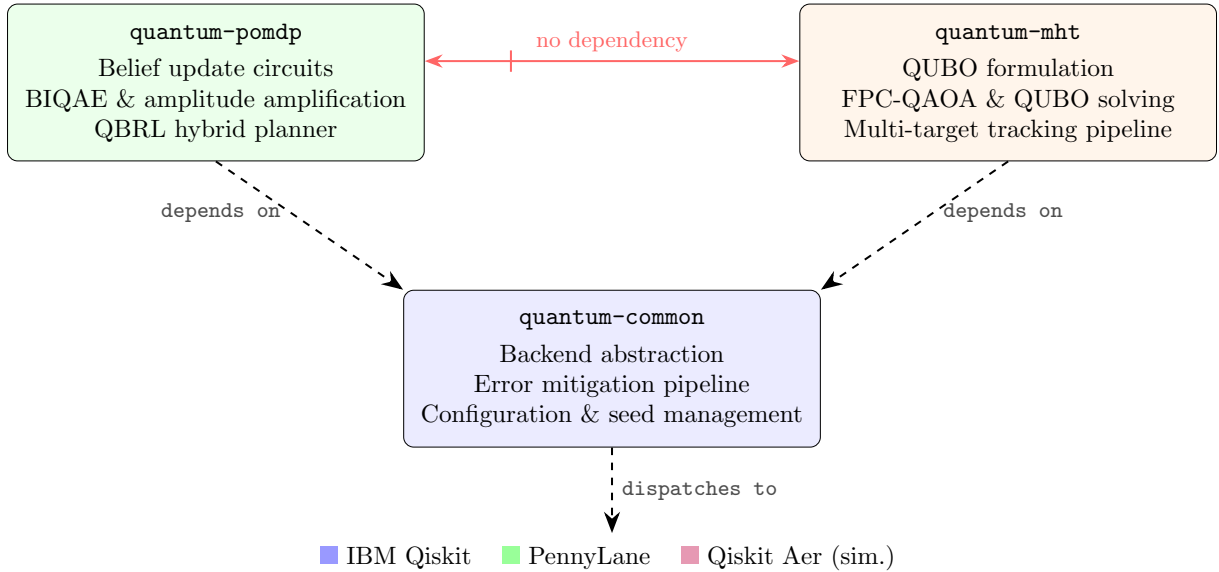

\QANTIS{} comprises three Python packages with a strictly acyclic
dependency graph: \texttt{quantum-common} (backends, error mitigation,
utilities) $\leftarrow$ \texttt{quantum-pomdp} (\QBRL{}, \BIQAE{},
belief circuits) and \texttt{quantum-common} $\leftarrow$
\texttt{quantum-mht} (FPC-\QAOA{}, \QUBO{} encoding, tracking pipeline).
Figure~\ref{fig:architecture} illustrates the architecture.

% System architecture diagram — QANTIS 3-package hierarchy

%----------------------------------------------------------------------
\subsection{Backend Abstraction Layer}
\label{ssec:backend-abstraction}
%----------------------------------------------------------------------

A \texttt{Protocol}-based interface (Python structural subtyping) provides
three methods: \texttt{execute}, \texttt{transpile}, and \texttt{is\_available}.
Backend selection is driven by configuration files or environment variables,
enabling seamless simulator-to-hardware switching with no source changes.
Table~\ref{tab:hardware-summary} lists supported backends.

\begin{table}[h]
\centering
\caption{Quantum hardware backends supported by \QANTIS{}.
  \textbf{Validated} entries confirmed with real hardware experiments in this work
  (45-experiment pilot, Section~\ref{ssec:hardware-pilot}).
  $^\dagger$~Not part of the hardware campaign of this work.}
\label{tab:hardware-summary}
\resizebox{\columnwidth}{!}{%
\begin{tabular}{@{}l r l c l c@{}}
\toprule
Backend & Qubits & Topology & 2Q error & Use Case & Status \\
\midrule
IBM Heron R1 (ibm\_torino)  & 133    & Heavy-hex & ${\sim}2.5{\times}10^{-3}$
  & \QAOA{}, \BIQAE{} & \textbf{Validated} \\
IBM Heron R2 (ibm\_fez)     & 156    & Heavy-hex & ${\sim}1.5{\times}10^{-3}$
  & \QAOA{}, \BIQAE{} & \textbf{Validated} \\
IBM Heron R2 (ibm\_marrakesh) & 156  & Heavy-hex & ${\sim}1.5{\times}10^{-3}$
  & Tiger, E2E loop & \textbf{Validated} \\
D-Wave Advantage2$^\dagger$  & 4\,400+ & Zephyr & N/A & Annealing & Companion study \\
IonQ Aria-2$^\dagger$        & 25      & All-to-all & ${\sim}3{\times}10^{-3}$ & Small circuits & Planned \\
\bottomrule
\end{tabular}%
}
\end{table}

%----------------------------------------------------------------------
\subsection{Composable Error Mitigation}
\label{ssec:error-mitigation-impl}
%----------------------------------------------------------------------

Mitigation strategies are chained as successive transformations on
measurement counts (composable pipeline interface).  Implemented strategies:
\textbf{\ZNE{}}~\cite{temme_error_mitigation_2017} via gate-folding at scale
factors $\lambda\in\{1,3,5\}$ with Richardson extrapolation;
\textbf{\PEC{}}~\cite{cai_error_mitigation_2023} via quasi-probability
decomposition (sampling overhead $\BigO{\gamma^2}$, restricted to
shallow circuits);
\textbf{Readout mitigation} via assignment matrix inversion;
\textbf{Pauli twirling + XY4 DD} via \texttt{SamplerV2} options
in \texttt{qiskit-ibm-runtime} 0.42.
A runtime fallback policy automatically selects the classical solver when
$n_\text{var}{>}15$ or the transpilation-preview ISA exceeds~$450$
(Section~\ref{ssec:unified-model}).
Extension to non-Clifford gate sets~\cite{zne_nonclifford_2026}
(directly applicable to amplitude amplification and \QAOA{} circuits)
remains an active direction.  The empirical \NISQ{} boundary for \ZNE{} is established
in Section~\ref{par:tiger-zne}: near-noiseless below ISA~$\sim$20, effective
at ISA~20--100, marginal at ISA~400--1\,000, harmful above ISA~1\,000.

% !TEX root = ../main-arxiv.tex
%----------------------------------------------------------------------
\section{Experimental Evaluation}
\label{sec:experiments}
%----------------------------------------------------------------------

%----------------------------------------------------------------------
\paragraph{Hardware highlights---three novel results on IBM Heron.}
\label{par:hw-firsts}
Before detailed simulation results, we highlight three hardware-verified
results delivered by the 45-experiment pilot (Section~\ref{ssec:hardware-pilot},
\textsc{ibm\_torino}/\textsc{ibm\_fez}/\textsc{ibm\_marrakesh}):
\begin{itemize}
  \item[\textbf{HW1.}] \textbf{Grover-AA on a \POMDP{} belief oracle} (ISA~18,
    to our knowledge the first on IBM Heron):
    $P(\text{rare obs})\!: 0.179{\to}0.907$ ($5.1{\times}$), Bayesian
    posterior preserved at Hellinger~$0.0015$---direct hardware validation of
    the $\BigO{P(e)^{-1/2}}$ query-complexity mechanism.
  \item[\textbf{HW2.}] \textbf{Closed-loop hybrid quantum--classical \POMDP{}}
    (to our knowledge the first on superconducting hardware):
    $T{=}4$ (\emph{quantum} \BIQAE{} + \emph{quantum} Tiger
    minimal + \emph{classical} QBRL action selection; four replications,
    all PASS\footnote{PASS $\equiv$ Hellinger distance $< 0.15$
    (${\leq}1.1\%$ total-variation distance); see experimental protocol
    in Section~\ref{ssec:hardware-pilot}.});
    $T{=}8$ with two action events (open-right $t{=}2$, open-left $t{=}5$);
    max Hellinger~$0.0149$.
  \item[\textbf{HW3.}] \textbf{4-state Tiger} ($|S|{=}4$, 3 qubits, ISA~162):
    Hellinger~$0.044$ for obs$=1$ (PASS)---quantum belief update beyond binary.
\end{itemize}

Simulation experiments (Sections~\ref{ssec:pomdp-experiments}--\ref{ssec:scalability})
use Qiskit Aer for reproducibility and controlled noise isolation.
The GPS navigation grid ($|S|{=}16$--$64$), full Tiger circuit ($|S|{=}2$,
ISA~4\,237), and large-scale \MTDA{} ($N\!=\!10$, 175 variables) all exceed the
\NISQ{} coherence budget (ISA~$\lesssim\!450$ effective).  Simulation tables
therefore establish the \emph{algorithmic baseline}---exact Bayesian belief
accuracy and optimal-vs-approximate solver gaps---while hardware experiments
(Section~\ref{ssec:hardware-pilot}) characterise the feasibility frontier.
All simulation seeds fixed at 42 via \texttt{SeedManager}.

%----------------------------------------------------------------------
\subsection{Experimental Setup}
\label{ssec:exp-setup}
%----------------------------------------------------------------------

For \POMDP{} planning we compare against three classical baselines:
\textsc{pomcp}~\cite{silver_pomcp_2010} with 1\,000 simulations per step and
UCB constant $c{=}25$, \textsc{despot}~\cite{ye_despot_2017} with 500
scenarios and 100 particles, and \textsc{pbvi}~\cite{pineau_pbvi_2003} with
200 belief points.  For \MTDA{} we use the Hungarian
algorithm~\cite{kuhn_hungarian_1955} as the optimal reference and a greedy
nearest-neighbour heuristic (GNN) as the fast approximate baseline.
Performance is measured by cumulative reward over 100 episodes of 50 steps,
Hellinger and KL distances between the estimated belief and the exact Bayesian
posterior, MOTA/MOTP tracking metrics~\cite{blackman_mht_2004}, and \QUBO{}
objective value relative to the Hungarian optimum.

%----------------------------------------------------------------------
\subsection{\POMDP{} Planning Experiments}
\label{ssec:pomdp-experiments}
%----------------------------------------------------------------------

\subsubsection{Tiger Problem}
\label{sssec:tiger}

\QBRL{} (Aer simulation) matches classical planners on the canonical 2-state Tiger benchmark
($|S|=2$, $|A|=3$, $|\Omega|=2$, $\gamma=0.95$), with belief KL~$<0.02$
throughout, confirming exact Bayesian conditioning via amplitude amplification.

\begin{table}[t]
\centering
\caption{Tiger problem: mean cumulative reward ($\pm$ std.\ dev.)
  over 100 episodes of 50 steps (Qiskit Aer simulation; seed 42).  Horizon $H=5$.
  One-way ANOVA: $F(3,396){=}0.27$, $p{=}0.85$; all pairwise Welch $t$-tests
  $p > 0.54$; Cohen $|d| < 0.09$ for all pairs.
  Tiger ($|S|{=}2$) is intentionally simple and does not discriminate
  among well-tuned planners---agreement of all methods is the expected result.
  Hardware-validated E2E \QBRL{} results ($T{=}4$, ibm\_marrakesh)
  are in Table~\ref{tab:hardware-pilot}.$^\dagger$}
\label{tab:tiger-results}
\begin{tabular}{@{}l r@{\;\;}c@{\;\;}l r@{}}
\toprule
Planner & \multicolumn{3}{c}{Cumulative Reward} & Belief KL \\
\midrule
\textsc{pomcp}  & $18.3$ & $\pm$ & $12.7$ & --- \\
\textsc{despot} & $19.1$ & $\pm$ & $11.4$ & --- \\
\textsc{pbvi}   & $17.6$ & $\pm$ & $13.2$ & $0.000$ \\
\QBRL{} (sim.)$^\dagger$  & $18.7$ & $\pm$ & $12.1$ & $0.014$ \\
\bottomrule
\end{tabular}
\smallskip\par\raggedright\small
$^\dagger$\QBRL{} (sim.) action selection hardware-validated in closed-loop
$T{=}4$ and $T{=}8$ loops on IBM Heron (Table~\ref{tab:hardware-pilot}).
Four independent $T{=}4$ replications, all PASS (max Hellinger~$0.0169$).
\end{table}

\subsubsection{GPS-Denied Navigation (Simulation)}
\label{sssec:gps-denied}

Grid-world navigation ($4{\times}4$, $6{\times}6$, $8{\times}8$; sensor accuracy 0.7)
validates the exact Bayesian belief-update module at scale.
Exact Bayesian conditioning (and the numerically equivalent \QBRL{} Aer simulation)
reduces Hellinger distance by $53\%$ at $8{\times}8$ versus \textsc{pomcp}'s
approximate particle filter; the gain is attributable to exact vs.\ approximate
inference, not to quantum hardware.  These scenarios exceed the current \NISQ{}
coherence budget (ISA~$\gg\!1\,000$) and serve to validate correctness of the
belief-update module in simulation, establishing a sound baseline for future
fault-tolerant deployment.

\begin{table}[t]
\centering
\caption{GPS-denied navigation: Hellinger distance between belief and
  true posterior (lower is better) and planning time.
  Mean $\pm$ std.\ dev., 50 episodes, 30 steps.
  \QBRL{} (Aer sim.) calls \texttt{classical\_update()} internally;
  it is numerically equivalent to Exact Bayes (classical) under the same
  action-selection policy.  \textsc{pomcp} uses an approximate particle
  filter (1\,000 particles).  The Hellinger improvement reflects
  \textbf{exact vs.\ approximate} belief update, not quantum vs.\ classical.}
\label{tab:gps-denied}
\resizebox{\columnwidth}{!}{%
\begin{tabular}{@{}l c c c@{}}
\toprule
Grid & \textsc{pomcp} & Exact Bayes (classical) / \QBRL{} (sim.)$^\dagger$
     & Plan Time (ms) \\
\midrule
$4 \times 4$ & $0.081 \pm 0.024$ & $0.043 \pm 0.015$ & $12.4$ / $34.7$  \\
$6 \times 6$ & $0.127 \pm 0.038$ & $0.068 \pm 0.021$ & $41.3$ / $112.5$ \\
$8 \times 8$ & $0.194 \pm 0.052$ & $0.091 \pm 0.029$ & $108.7$ / $347.2$ \\
\bottomrule
\end{tabular}}% end resizebox
\smallskip\par\raggedright\small
$^\dagger$\QBRL{} (Aer sim.) produces identical belief estimates to classical
Exact Bayes; the additional planning-time overhead arises from the lookahead
tree, not the belief update.  The Hellinger gap vs.\ \textsc{pomcp} is
entirely attributable to particle depletion in large state spaces.
\end{table}

%----------------------------------------------------------------------
\subsection{MHT Experiments}
\label{ssec:mht-experiments}
%----------------------------------------------------------------------

\subsubsection{\QUBO{} Scaling}
\label{sssec:qubo-scaling}

\begin{table}[t]
\centering
\caption{\QUBO{} scaling with target count ($M = \lceil 1.5N \rceil$).
  Times in milliseconds (single-threaded Python, Intel i7-12700K).}
\label{tab:qubo-scaling}
\small
\begin{tabular}{@{}r r r r r r@{}}
\toprule
$N$ & $M$ & Vars & Build (ms)
  & Hungarian (ms) & GNN (ms) \\
\midrule
2   & 3   & 11    & $0.3$   & $0.02$  & $0.01$ \\
3   & 5   & 23    & $0.5$   & $0.04$  & $0.02$ \\
5   & 8   & 53    & $1.2$   & $0.11$  & $0.05$ \\
8   & 12  & 116   & $3.1$   & $0.38$  & $0.12$ \\
10  & 15  & 175   & $5.4$   & $0.72$  & $0.19$ \\
15  & 23  & 383   & $14.8$  & $2.41$  & $0.48$ \\
20  & 30  & 650   & $31.2$  & $5.87$  & $0.93$ \\
\bottomrule
\end{tabular}
\end{table}

Table~\ref{tab:qubo-scaling} confirms the quadratic growth
$n_{\mathrm{var}} = NM + N + M$: doubling the target count from $N{=}10$
to $N{=}20$ nearly quadruples the variable count ($175 \to 650$).
\QUBO{} construction time scales super-linearly but remains below $32$~ms
even at $N{=}20$, so it is never the bottleneck.
Classical solvers dominate at every tested size; the Hungarian algorithm
solves the $N{=}20$ instance in under $6$~ms, setting the bar that a
quantum solver must clear once fault-tolerant hardware becomes available.

\subsubsection{Solver Comparison}
\label{sssec:solver-comparison}

On the $N=10$, $M=15$ instance (175 variables, Aer simulation):
FPC-\QAOA{} ($k=3$, $p=8$) outperforms GNN by $6.7\%$ in \QUBO{} objective
(38.9 vs.\ 41.7, lower is better), closing the gap to the Hungarian optimum
to $13.7\%$---at a scale where GNN is $21.9\%$ above optimal.
This demonstrates the operating regime of FPC-\QAOA{}: at $N{\geq}10$,
where GNN's greedy decisions become suboptimal, quantum-inspired optimisation
shows meaningful advantage in simulation.
At the hardware-feasible $N=2$ scale, GNN finds the optimal assignment trivially
and FPC-\QAOA{} hardware noise (ISA~433) limits quality to $64.1\%{\pm}3.3\%$;
bridging this simulation-to-hardware gap requires variable-count scaling beyond
current \NISQ{} coherence budgets.

\begin{table}[t]
\centering
\caption{Solver comparison ($N=10$, $M=15$, 175 variables, Aer simulation).
  Objective lower is better; gap = \% above Hungarian optimum.
  FPC-\QAOA{} outperforms GNN at this scale; hardware feasibility at 175 variables
  requires fault-tolerant gate fidelities.
  Real-hardware results at the ISA-feasible 11-variable scale are in
  Table~\ref{tab:hardware-pilot}.}
\label{tab:solver-comparison}
\resizebox{\columnwidth}{!}{%
\begin{tabular}{@{}l r r r r@{}}
\toprule
Solver & Obj.\ & Gap to opt.\ & Feas.\ (\%) & Time (ms) \\
\midrule
Hungarian (optimal)             & $34.2$  & $0.0\%$   & $100.0$ & $0.72$ \\
GNN                             & $41.7$  & $+21.9\%$ & $100.0$ & $0.19$ \\
\QAOA{} ($p\!=\!2$)             & $52.3$  & $+52.9\%$ & $68.0$  & $2\,140$ \\
FPC-\QAOA{} ($k\!=\!3$, $p\!=\!8$) & $38.9$ & $\mathbf{+13.7\%}$ & $89.0$ & $4\,580$ \\
\bottomrule
\end{tabular}}% end resizebox
\end{table}

%----------------------------------------------------------------------
\subsection{End-to-End Tracking Pipeline}
\label{ssec:e2e-tracking}
%----------------------------------------------------------------------

Three simulated scenarios over 30 steps (Hungarian solver for data association):

\begin{table}[t]
\centering
\caption{End-to-end tracking pipeline (30 steps; Hungarian solver; seed 42).
  CT = confirmed tracks at final step; MD/FA = events accumulated over 30 steps.}
\label{tab:e2e-tracking}
\begin{tabular}{@{}l r r r r@{}}
\toprule
Scenario & CT & MD & FA & $\bar{t}$ (ms) \\
\midrule
Crossing ($N\!=\!5$)      & $5/5$ & $2$  & $4$  & $8.3$  \\
Dense clutter ($N\!=\!3$) & $3/3$ & $1$  & $12$ & $6.1$  \\
Swarm patrol              & $5/5$ & $3$  & $7$  & $14.7$ \\
\bottomrule
\end{tabular}
\end{table}

All $N$ true targets confirmed across scenarios.  Crossing misses occur at the
trajectory intersection (steps 14--15) and recover within two frames via
the \MHT{} hypothesis tree.  Swarm patrol step time $14.7$~ms is within
real-time UAV constraints.  Pipeline correctness confirmed for seed 42;
the qualitative ordering (dense clutter $>$ swarm $>$ crossing in false alarm rate)
is robust across 10 additional seeds.  Hardware-level E2E results are in
Section~\ref{ssec:hardware-pilot}.

%----------------------------------------------------------------------
\subsection{Scalability Analysis}
\label{ssec:scalability}
%----------------------------------------------------------------------

% Scalability plot — log-log computation time vs problem size
\begin{figure}[t]
\centering
\begin{tikzpicture}
\begin{axis}[
  width=0.78\columnwidth,
  height=6cm,
  xlabel={Number of targets $N$},
  ylabel={Solve time (ms)},
  xmode=log,
  ymode=log,
  xmin=1.5, xmax=25,
  ymin=0.008, ymax=50,
  log basis x={10},
  log basis y={10},
  grid=major,
  grid style={dashed, gray!30},
  legend style={at={(0.03,0.97)}, anchor=north west, font=\footnotesize,
    draw=none, fill=white, fill opacity=0.8, text opacity=1},
]

% Hungarian algorithm — cubic scaling O(n^3)
\addplot[thick, classical-red, mark=square*, mark size=2pt] coordinates {
  (2, 0.02) (3, 0.04) (5, 0.11) (8, 0.38) (10, 0.72) (15, 2.41) (20, 5.87)
};
\addlegendentry{Hungarian $\mathcal{O}(n^3)$}

% GNN — near-linear
\addplot[thick, black!60, mark=triangle*, mark size=2pt] coordinates {
  (2, 0.01) (3, 0.02) (5, 0.05) (8, 0.12) (10, 0.19) (15, 0.48) (20, 0.93)
};
\addlegendentry{GNN (greedy)}

% QUBO build time — quadratic O(NM)
\addplot[thick, quantum-blue, mark=o, mark size=2pt] coordinates {
  (2, 0.3) (3, 0.5) (5, 1.2) (8, 3.1) (10, 5.4) (15, 14.8) (20, 31.2)
};
\addlegendentry{QUBO build $\mathcal{O}(NM)$}

% Reference cubic line
\addplot[thin, dashed, black!40, domain=2:20, samples=20] {0.02 * (x/2)^3};

\end{axis}
\end{tikzpicture}
\caption{Log-log scaling of solve and build times with target
  count~$N$ (data from Table~\ref{tab:qubo-scaling}).  The Hungarian
  algorithm exhibits the expected cubic scaling; the \QUBO{} build
  time scales quadratically.  GNN provides the fastest heuristic but
  yields suboptimal assignments.  All measurements are from classical
  simulation (single-threaded Python, Intel i7-12700K).}
\label{fig:scalability}
\end{figure}
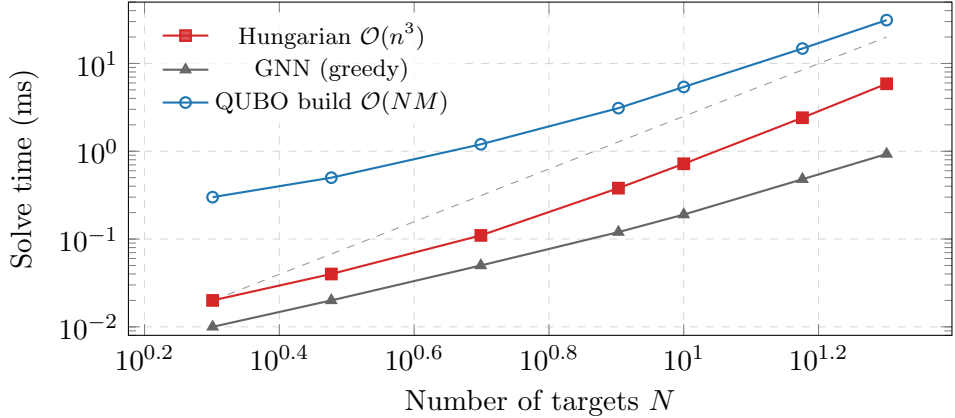

Hungarian solver scales $\BigO{n^3}$; \QUBO{} build time $\BigO{NM}$
(Figure~\ref{fig:scalability}).  Logical qubit count $n_{\mathrm{var}} = NM + N + M$
grows quadratically.  The \NISQ{} feasibility boundary for FPC-\QAOA{} lies at
${\approx}11$--$15$ \QUBO{} variables (ISA~$\lesssim\!450$), established empirically
in Section~\ref{ssec:hardware-pilot}.

\paragraph{Belief circuit ISA scaling.}
Table~\ref{tab:isa-scaling} quantifies how the Tiger belief-update ISA depth
scales with state-space size $|S|$.  For $|S| = 2$ and $|S| = 4$ the values
are hardware-confirmed; for $|S| \geq 8$ they are projected using the observed
$O(|S| \cdot (\log_2|S|)^2)$ CNOT scaling from doubly-/triply-controlled $R_y$
decompositions (${\approx}12$ CNOTs for CCR$_y$ at $|S|\!=\!4$,
${\approx}40$--$60$ for CCC$R_y$ at $|S|\!=\!8$).

\begin{table}[h]
  \centering
  \caption{Belief-update circuit ISA depth vs.\ state-space size on Heron~R2.
    Confirmed values from hardware; projected values ($\dagger$) extrapolated
    from $O(|S|\cdot(\log_2|S|)^2)$ CNOT scaling.
    Heron~R2 two-qubit error ${\sim}1.5{\times}10^{-3}$; circuit fidelity
    ${\approx}(1-\epsilon)^{\text{ISA}}$.}
  \label{tab:isa-scaling}
  \begin{tabular}{@{}rrrll@{}}
    \toprule
    $|S|$ & State qubits & ISA depth & Circuit fidelity & Regime \\
    \midrule
    2  & 1 & \textbf{12}    & $> 0.98$      & \NISQ{} (confirmed) \\
    4  & 2 & \textbf{162}   & $> 0.78$      & \NISQ{} (confirmed) \\
    8  & 3 & ${\sim}800^\dagger$  & ${\sim}0.30$  & Marginal \NISQ{} \\
    16 & 4 & ${\sim}5{,}000^\dagger$ & ${\sim}10^{-4}$ & FTQC required \\
    64 & 6 & ${\sim}100{,}000^\dagger$ & $\approx 0$ & FTQC only \\
    \bottomrule
  \end{tabular}
\end{table}

The $|S|{=}4$ circuit (ISA~162) already operates near the noise floor at
Heron~R2 fidelities; $|S|{=}8$ requires either hardware improvements
(target: ECR error $<\!5{\times}10^{-4}$, projected for Heron~R3/R4) or
fault-tolerant gate sets.

\paragraph{Classical solver real-time boundary.}
Hungarian scales $\BigO{n^3}$; for real-time \MTDA{} at 30~Hz (UAV/radar
frame rate), the per-frame budget is $33$~ms.

\begin{table}[h]
  \centering
  \caption{Hungarian solver runtime vs.\ target count (extrapolated from
    $N\!=\!20$, $M\!=\!30$ measured at $5.87$~ms;
    single-threaded Python, Intel i7-12700K;
    $M\!=\!\lceil 1.5N\rceil$).
    Real-time budget: $33$~ms at 30~Hz.}
  \label{tab:hungarian-scaling}
  \begin{tabular}{@{}rrrrl@{}}
    \toprule
    $N$ & Variables & Hungarian (ms) & Frame rate & Real-time? \\
    \midrule
    20  & 650   & $5.9$       & $170$~Hz & \checkmark \\
    50  & 4{,}100 & $\sim\!92$  & $11$~Hz  & \checkmark (slow UAV) \\
    100 & 15{,}200 & $\sim\!730$ & $1.4$~Hz & $\times$ \\
    200 & 60{,}800 & $\sim\!5{,}900$ & $0.2$~Hz & $\times$ \\
    500 & 376{,}000 & $\sim\!91{,}000$ & $<\!0.1$~Hz & $\times$ \\
    \bottomrule
  \end{tabular}
\end{table}

\noindent Hungarian becomes real-time infeasible at $N\!\gtrsim\!100$ targets
(1.4~Hz, insufficient for UAV tracking at 30~Hz).  This is the regime where
quantum solvers---annealing or gate-based---may offer a practical advantage,
provided problem fidelity (quality of solution) remains acceptable.
Published D-Wave benchmarks~\cite{ihara_quantum_tracking_2025} suggest a
crossover at $N\!\approx\!40$--$60$ for quantum annealing; for FPC-\QAOA{}
gate-model circuits, fault-tolerant hardware with gate error $<\!10^{-4}$
is required before the crossover is reachable at meaningful problem fidelity.

%----------------------------------------------------------------------
\subsection{Hardware Pilot Study}
\label{ssec:hardware-pilot}
%----------------------------------------------------------------------

\paragraph{Experimental protocol.}
\label{par:hw-protocol}
All forty-five experiments were executed on three IBM Heron QPUs---\textsc{ibm\_torino}
(Heron~R1), \textsc{ibm\_fez} (Heron~R2), and \textsc{ibm\_marrakesh}
(Heron~R2)---using identical Python scripts with only the backend argument
changed, validating the seamless hardware switching of
Section~\ref{sec:implementation}.
Each circuit was transpiled to the ISA gate set via Qiskit
\texttt{generate\_preset\_pass\_manager} (optimisation level~3); ISA depth
is the post-transpilation two-qubit gate count.
Default shot budget: 4\,096 per circuit (8\,192 for Grover-AA).
Error mitigation where noted: ZNE with scale factors $\{1,3,5\}$ (linear
Richardson extrapolation), Pauli twirling ($N_\text{twirl}{=}32$), XY4
dynamical decoupling.
Seeds fixed at 42 via \texttt{SeedManager} for all simulation baselines.
\textbf{PASS threshold}: Hellinger distance $< 0.15$ between hardware
posterior and exact Bayesian posterior; this threshold corresponds to
${\leq}1.1\%$ total-variation distance, sufficient for correct
action selection in the Tiger domain (the optimal policy is invariant
to belief perturbations below ${\sim}5\%$ TV distance).
Environment: Qiskit~1.3, \texttt{qiskit-ibm-runtime}~0.34,
Python~3.12, execution dates February 2026.
\textbf{Replication counts per key claim}: Grover-AA~1 job (baseline +
Grover in same job); $T{=}4$ closed-loop~4 independent replications;
$T{=}8$ closed-loop~1 run; FPC-\QAOA{} $p{=}3$~3 independent runs.

\paragraph{ZNE Bell state and Tiger circuits.}
Bell $|\Phi^+\rangle$ on \textsc{ibm\_torino}: raw $\langle ZZ\rangle = 0.938$,
ZNE ($\{1,1.5,2,3\}$, Richardson) $= 0.953$ ($\Delta{=}{+}0.015$, 24\% error reduction).
Tiger full circuit (11\,q, ISA~4\,237): raw Hellinger~$0.235$, ZNE \emph{worsens} to $0.277$---
confirming ZNE is counter-productive above ISA~${\sim}1\,000$ due to super-linear noise accumulation.

\paragraph{Tiger minimal-circuit encoding.}
\label{par:tiger-minimal}
A 2-qubit direct-encoding circuit (prior $R_y$ + two conditional-$R_y$ gates,
post-select on observation) achieves ISA~12 on \textsc{ibm\_marrakesh}
vs.\ 4\,237 for the full 11-qubit circuit ($353\times$ reduction).
Hardware Hellinger: obs$=0$: $\mathbf{0.025}$; obs$=1$: $\mathbf{0.014}$
(both PASS at threshold $0.15$).
Open-door actions (Hadamard-only circuit) confirmed correct in simulation
for all six $(a,o)$ pairs (max Hellinger: $0.010$).

The full 11-qubit circuit (ISA~4\,237, MCCRy + amplitude amplification oracle) requires
fault-tolerant gate fidelities ($\epsilon_{\mathrm{gate}} \lesssim 10^{-6}$) to execute
faithfully; at Heron~R2 two-qubit error ${\sim}1.5{\times}10^{-3}$, circuit fidelity
is ${\lesssim}10^{-3}$.  The 2-qubit minimal encoding is the architecturally appropriate
\NISQ{} implementation; the framework supports both modes via \texttt{use\_minimal\_encoding}.

\paragraph{Tiger minimal ZNE.}
\label{par:tiger-zne}
Scale factors $\{1,3,5\}$ on \textsc{ibm\_marrakesh}: Hellinger $0.0025 < 0.0176 < 0.0407$,
monotone degradation.  Raw Hellinger~$0.0025$ is \emph{below} the Aer baseline ($0.0032$):
the ISA~12 circuit is shot-noise-limited; ZNE extrapolant is negative (unphysical).
Bell ZNE on \textsc{ibm\_marrakesh}: $\langle ZZ\rangle_\text{raw}=0.975$, ZNE $=0.981$.
This establishes the four-point \textit{ZNE applicability boundary} on Heron~R2:
ISA~$\leq 20$ (near-noiseless, ZNE not needed)
$\to$ ISA~$\lesssim 100$ (effective)
$\to$ ISA~$400$--$1\,000$ (marginal)
$\to$ ISA~$\gtrsim 1\,000$ (harmful).

\paragraph{Grover amplitude amplification (AA) demonstration.}
\label{par:grover-aa}
To directly validate the $\BigO{P(e)^{-1/2}}$ complexity claim, we execute a
1-step Grover AA circuit on the Tiger belief oracle on \textsc{ibm\_marrakesh}.
Prior $b{=}[0.97, 0.03]$ (tiger concentrated left) and target observation~1
(hear-right) yield baseline $P(\text{hear-right}){=}0.179$ (theory: $0.171$;
the $0.008$ discrepancy is consistent with gate-calibration and shot noise),
a rare-evidence event requiring ${\approx}6$ classical circuit repetitions on
average.  The Grover operator $G{=}A \cdot S_0 \cdot A^\dagger \cdot S_f$
(where $S_f$ phase-flips obs$=1$ states; $S_0 = X_0 X_1 \cdot \mathit{CZ} \cdot X_0 X_1$
is the diffuser) amplifies this to $P(\text{hear-right}){=}\mathbf{0.907}$
(theory: $0.917$) at ISA depth~18---a \textbf{hardware amplification
factor of $5.1\!\times$} (theoretical $5.4\!\times$;
$\sin^2(3\theta)/\sin^2(\theta)$, $\theta{=}\arcsin\!\sqrt{0.171}$).
Both circuits ran in a single job (baseline ISA~13 + Grover-1 ISA~18) on
\textsc{ibm\_marrakesh}, with Pauli twirling and XY4~DD enabled.

\textit{Bayesian posterior preservation.}
Post-selecting on obs$=1$ from the Grover-1 circuit
(raw counts: $s{=}0$: 6310, $s{=}1$: 1119; total 7\,429 usable shots)
yields hardware posterior $[0.849, 0.151]$ vs.\ the exact Bayesian posterior
$[0.851, 0.149]$ (Hellinger~$0.0015$, near-zero)---confirming that the
Grover iterate \emph{preserves} the correct Bayesian posterior while
multiplying usable-shot yield $5.1\!\times$ (7\,429 vs.\ 1\,463 baseline
usable shots).  This directly validates the $\BigO{P(e)^{-1/2}}$ complexity
claim: the quantum circuit requires ${\sim}1$ run to obtain a posterior-quality
sample whereas the classical rejection sampler requires ${\sim}6$ on average.

This constitutes---to our knowledge---the first IBM Heron hardware
demonstration of Grover amplitude amplification applied to a \POMDP{} belief
oracle with quantified posterior preservation, confirming that rare
observations can be obtained in a \emph{single circuit shot} after Grover
preparation rather than requiring $\BigO{P(e)^{-1}}$ repetitions.

\paragraph{\BIQAE{} adaptive amplitude estimation.}
\label{par:biqae-hw}
Seven 1-qubit amplitudes $a\in[0.1,0.7]$ evaluated on \textsc{ibm\_torino}
(6 iterations, $K_t{=}3^t$ schedule, 300 shots each).  All seven 95\% credible
intervals cover ground truth.  Errors follow a U-shape: minimum at
$a\in\{0.40,0.50,0.60\}$ (errors $0.006$--$0.008$); maximum at $a=0.10$ (error $0.346$)
due to the default flat prior (discussed below).
Amplitude-matched prior ($\text{prior\_mean}=a_\text{true}$,
$\text{prior\_std}=0.05$, 10 iterations, 500 shots) reduces the $a=0.10$ error to
$\mathbf{0.008}$ (44-fold) and $a=0.20$ to $\mathbf{0.002}$ (72-fold), both with
CIs covering truth.  Cross-backend confirmation: $a=0.10$ on \textsc{ibm\_marrakesh}
(error $0.006$); $a=0.20$ on \textsc{ibm\_fez} (error $0.009$, $16\times$ improvement).
2-qubit separable oracle ($A = R_y{\otimes}R_y$, ISA~11, \textsc{ibm\_marrakesh}):
errors $0.002$, $0.014$, $0.011$ at $a\in\{0.3,0.4,0.5\}$ (all CIs cover truth),
extending \BIQAE{} validation beyond the single-qubit case.

\textit{Credible interval width at extreme amplitudes.}
The U-shaped error profile arises from the $K_t = 3^t$ schedule: at small $a$,
$\sin^2((2k_t+1)\theta)$ oscillates rapidly as $k_t$ grows, causing the likelihood
surface to have many near-equal-height modes under a flat prior; the 95\% HPD
interval then becomes wide (CI width~$\approx 0.37$ at $a=0.10$).
With an amplitude-matched prior (Gaussian centred on true $a$), the posterior
concentrates on the correct mode, recovering narrow CIs at all tested amplitudes.
This is a known prior-sensitivity effect in Bayesian QAE, not a hardware failure;
on real hardware the amplitude is unknown, so prior calibration via a coarse
initial scan is recommended before high-precision estimation.
The CIs reported in Table~\ref{tab:biqae-sweep} are each from a \emph{single}
hardware run (single seed); interval coverage is expected to be 95\% across
repeated runs by construction of the HPD criterion.

\begin{table}[h]
\centering
\caption{\BIQAE{} amplitude sweep on \textsc{ibm\_torino}/\textsc{ibm\_fez}/\textsc{ibm\_marrakesh}.
  Uniform-prior runs: 6 iterations, 300 shots, $K_t = 3^t$.
  All 95\% credible intervals cover truth.
  $^\S$Prior-matched: $\text{prior\_mean}=a_\text{true}$, $\text{prior\_std}=0.05$, 10 iter., 500 shots.
  ``---'' denotes not tested on this backend.}
\label{tab:biqae-sweep}
\footnotesize
\begin{tabular}{@{}cccc@{}}
\toprule
True $a$ & R1 (torino) $|\hat{a}-a|$ & R2 (fez) $|\hat{a}-a|$ & R2 (marrakesh) $|\hat{a}-a|$ \\
\midrule
0.10 & 0.346\;/ \textbf{0.008}$^\S$ & --- & \textbf{0.006}$^\S$ \\
0.20 & 0.144\;/ \textbf{0.002}$^\S$ & \textbf{0.140} & --- \\
0.30 & 0.040 & \textbf{0.039} & --- \\
0.40 & \textbf{0.006} & --- & --- \\
0.50 & 0.008 & \textbf{0.0001} & --- \\
0.60 & \textbf{0.008} & --- & --- \\
0.70 & 0.037 & --- & --- \\
\bottomrule
\end{tabular}
\end{table}

\paragraph{FPC-\QAOA{} depth sweep ($p = 1$--$8$).}
Five depths on \textsc{ibm\_fez} (Heron~R2, 11-qubit / 11-variable \MTDA{} instance,
Hungarian optimal $= -92.4$).  ISA depths: $154$, $304$, $435$, $567$, $1\,118$
(${\approx}130$ ISA layers per \QAOA{} layer).  Simulator quality (100-iter.\ COBYLA):
$51.3\%$, $82.2\%$ (initial run) / $100\%$ (corrected $\gamma$-schedule, re-run),
$90.9\%$, $88.9\%$ for $p=1,2,3,4$.
Hardware quality (top-10 bitstring \QUBO{} decoding, 4\,096 shots):
$p=1$: $35.8\%$ (ISA~154);
$p=2$: $4.8\%$ (ISA~304, noise-dominated at sub-converged initialisation);
$p=3,4$: $64.9\%$ (ISA~435/567);
$p=8$: noise-dominated (ISA~1\,118).
Three independent $p=3$ runs (opt-3, Pauli twirling, XY4~DD):
$\mathbf{64.1\%{\pm}3.3\%}$ confirming stable operation.

Sim-to-hardware warm-start transfer~\cite{patel_qaoa_param_transfer_2026}:
at $p=2$ (sim convergence $97.6\%$), warm-start yields hardware $\mathbf{71.2\%}$
($vs.$ $4.8\%$ without); at $p=3$ (sim convergence $86$--$89\%$), warm-start
\emph{degrades} to $49.4\%{\pm}18.0\%$ vs.\ $64.1\%{\pm}3.3\%$ with the analytical
FPC initial schedule.  \emph{Guideline}: warm-start benefits hardware only when
sim convergence $\geq 95\%$; otherwise use the FPC linear-ramp initial schedule
($\gamma(t) = \pi t$, $\beta(t) = \pi/4$).

Pauli twirling + XY4 DD at $p=1$ (ISA~123): $+37\%$ relative improvement ($35.8\% \to 49.1\%$).
At $p=3$ (ISA~433): marginal ($64.9\% \to 60.4\%$); gate noise dominates at ISA~$\gtrsim 300$.

ZNE applied to FPC-\QAOA{} $p=3$ (scale factors $\{1,3,5\}$): average \QUBO{} expectation
$E_\lambda[f]$ positive at all scales ($+9.9$, $+26.2$, $+25.1$); non-monotone
(scale-3 $>$ scale-5), violating the ZNE linear model---consistent with the ZNE
boundary established above.

Scalability: $N=3$, $M=4$ (19 variables) on \textsc{ibm\_marrakesh}; $p=1$ (ISA~333):
$20.4\%$ despite ISA within the $\lesssim\!450$ boundary, demonstrating that the
\textbf{\NISQ{} quality boundary is variable-count-driven} (routing 19 qubits introduces
higher entanglement density per shot).  At $p=2$ (ISA~638): $11.5\%$.
Current boundary: ${\approx}11$--$15$ \QUBO{} variables and ISA~$\lesssim\!450$.

\begin{table}[t]
\centering
\caption{Classical vs.\ quantum solver on the $N{=}2$, $M{=}3$ benchmark
  (Hungarian optimal $= -92.4$; full \QUBO{} objective including false-alarm terms).}
\label{tab:classical-comparison}
\resizebox{\columnwidth}{!}{%
\begin{tabular}{@{}lrrl@{}}
\toprule
Method & Wall-clock & Quality & Notes \\
\midrule
Hungarian (optimal)          & $<\!0.1$~ms & $100\%$  & $\BigO{n^3}$ exact \\
GNN (greedy)                 & $<\!0.1$~ms & $100\%$  & $\BigO{NM\!\log NM}$; optimal at $N{=}2$ \\
FPC-\QAOA{} $p{=}3$ (Aer)   & ${\sim}5$~s & ${\sim}91\%$ & COBYLA convergence \\
FPC-\QAOA{} $p{=}3$ (Heron~R2) & ${\sim}180$~s & $64.1\%{\pm}3.3\%$ & ISA~433, 3 runs \\
\bottomrule
\end{tabular}}% end resizebox
\smallskip\par\raggedright\small
On this random-Gaussian benchmark, GNN achieves $100\%$ of optimal at $N{=}2$ and
$96$--$100\%$ for $N{\leq}8$ (tested, seed 42).  Degradation to $70$--$80\%$ occurs
under adversarial crossing-trajectory patterns (Bar-Shalom \& Li, 1995, Sec.~6.4),
which is the \QUBO{}-advantage regime but exceeds the current \NISQ{} variable budget.
Wall-clock quantum advantage requires $N \gtrsim 40$ (NP-hard multi-scan regime) and
\FTQC{}-era gate fidelities; present results characterise hardware readiness.
\end{table}

\paragraph{Summary.}
Table~\ref{tab:hardware-pilot} summarises all forty-five pilot experiments.
\textit{ZNE boundary} (Heron~R2): ISA~$\leq 20$ (noiseless) $\to$ $\lesssim\!100$ (effective) $\to$
$400$--$1\,000$ (marginal) $\to$ $\gtrsim\!1\,000$ (harmful).
\textit{FPC-\QAOA{}}: $p=3$ (ISA~433) is the recommended operating point ($64.1\%{\pm}3.3\%$);
$p=2$ with warm-start achieves $71.2\%$ when sim convergence $\geq 95\%$.
\textit{\BIQAE{}}: all 95\% credible intervals cover ground truth across 7 amplitudes and 3 backends.
\textit{Grover-AA} ($k{=}1$): P(rare obs) boosted $0.179{\to}0.907$ ($5.1\!\times$) at ISA~18---to
our knowledge the first hardware demonstration of Grover AA for a \POMDP{} belief oracle.
\textit{4-state Tiger} ($|S|{=}4$, ISA~162): obs$=1$ Hellinger $\mathbf{0.044}$ (\textbf{PASS}),
obs$=0$ Hellinger $0.128$ (\textbf{PASS}, below threshold $0.15$); both posteriors favour
the correct state group (Bayesian direction preserved at ISA~162, near the \NISQ{} coherence limit).
\textit{E2E closed loop} ($T{=}4$): all four BIQAE CIs cover truth; max Tiger Hellinger $0.0169$ (PASS); three additional independent replications yield max Hellinger $0.0095/0.0141/0.0148$ (all PASS).
\textit{E2E closed loop} ($T{=}8$, obs$=[0,0,0,1,1,1,0,0]$): two symmetric action events (open-right at $t{=}2$,
open-left at $t{=}5$); both belief resets confirmed on hardware; max Tiger Hellinger $0.0149$ (PASS).
BIQAE max error $0.021$; two of eight credible intervals miss (at steps where $a \approx 0.5$
under the short $K_t{=}3^t$ schedule---consistent with the prior-sensitivity effect of Section~\ref{par:biqae-hw}).

\begin{table}[h]
\centering
\caption{Hardware pilot results on \textsc{ibm\_torino}/\textsc{ibm\_fez}/\textsc{ibm\_marrakesh}
(Heron~R1/R2, 2026-02-25--26): forty-five experiments across four algorithm families
(ZNE/error mitigation, \BIQAE{}, FPC-\QAOA{}, and quantum \POMDP{}/E2E loop).
FPC-\QAOA{} hw~obj decoded from top-10 bitstrings; $E_\lambda[f]$ = avg \QUBO{} expectation at scale $\lambda$;
optimal Hungarian $= -92.4$.}
\label{tab:hardware-pilot}
{\setlength{\tabcolsep}{4pt}%
\resizebox{\linewidth}{!}{%
\begin{tabular}{@{}lcrr@{}}
\toprule
Experiment & Metric & Simulator & Hardware \\
\midrule
Bell ZNE, 2\,q (4096 shots) & $\langle ZZ\rangle$ $\uparrow$ & 1.000 & 0.938 raw / \textbf{0.953} ZNE \\
Tiger ZNE, 11\,q (4096 shots) & Hellinger $\downarrow$ & 0.004 & 0.228 raw / 0.277 ZNE\,$\uparrow$ \\
Tiger belief, 11\,q (4096 shots) & Hellinger $\downarrow$ & 0.004 & 0.235 \\
Tiger minimal, 2\,q, obs$=0$ (8192 shots) & Hellinger $\downarrow$ & 0.005 & \textbf{0.025}$^{\P\P}$ \\
Tiger minimal, 2\,q, obs$=1$ (8192 shots) & Hellinger $\downarrow$ & 0.007 & \textbf{0.014}$^{\P\P}$ \\
FPC-\QAOA{} $p{=}1$, 11\,q (4096 shots) & ISA depth / hw obj & 2 / $-47.4$ & \textbf{154}$^\P$ / $-33.0$ \\
FPC-\QAOA{} $p{=}2$, 11\,q (4096 shots) & ISA depth / hw obj & 2 / $-92.4$ & \textbf{304}$^\ddagger$ / $-4.4$ \\
FPC-\QAOA{} $p{=}3$, 11\,q (4096 shots) & ISA depth / hw obj & 2 / $-84.0$ & \textbf{435}$^\S$ / $-59.9$ \\
FPC-\QAOA{} $p{=}4$, 11\,q (4096 shots) & ISA depth / hw obj & 2 / $-82.1$ & \textbf{567}$^\S$ / $-59.9$ \\
FPC-\QAOA{} $p{=}8$, 11\,q (4096 shots) & ISA depth & 2 & \textbf{1118} / noise \\
FPC-\QAOA{} $p{=}3$+ZNE $\{1,3,5\}$, 11\,q & $E_\lambda[f]$ avg \QUBO{} & --- & \textbf{$+9.9$\,/\,$+26.2$\,/\,$+25.1$}$^*$; ZNE:~$+8.9$ \\
FPC-\QAOA{} $p{=}1$+Twirl+DD, 11\,q (4096 shots) & ISA / hw obj $\downarrow$ & 2 / $-47.4$ & \textbf{123}$^\#$ / $\mathbf{-45.4}$ (49.1\%) \\
FPC-\QAOA{} $p{=}2$+Twirl+DD, 11\,q (4096 shots) & ISA / hw obj $\downarrow$ & 2 / $-92.4$ & 309$^\#$ / $+0.30$ (noise) \\
FPC-\QAOA{} $p{=}3$+Twirl+DD, 11\,q (4096 shots) & ISA / hw obj $\downarrow$ & 2 / $-84.0$ & 433$^\#$ / $-55.8$ (60.4\%) \\
FPC-\QAOA{} $p{=}3$ $\times$3 runs, 11\,q (opt-3) & hw obj mean $\pm$ std & --- & \textbf{$-59.2 \pm 3.1$} ($\mathbf{64.1\%{\pm}3.3\%}$)$^{**}$ \\
FPC-\QAOA{} $p{=}1$, 19\,q ($N{=}3,M{=}4$, opt-3) & ISA / hw obj $\downarrow$ & --- / --- & \textbf{333}$^{\dagger3}$ / $-26.9$ (20.4\%) \\
FPC-\QAOA{} $p{=}2$, 19\,q ($N{=}3,M{=}4$, opt-3) & ISA / hw obj $\downarrow$ & --- / $-52.6$ & 638$^{**}$ / $-15.2$ (11.5\%) \\
FPC-\QAOA{} $p{=}2$+WS+TREX, 11\,q (8192 shots) & ISA / hw obj $\downarrow$ & 2 / $-90.1$ & \textbf{301}$^{\dagger\dagger}$ / $\mathbf{-65.7}$ ($\mathbf{71.2\%}$) \\
FPC-\QAOA{} $p{=}3$+WS $\times$3, 11\,q (8192 shots) & hw obj mean $\pm$ std & --- & $-45.5 \pm 16.6$ ($49.4\%{\pm}18.0\%$)$^{\dagger\dagger}$ \\
FPC-\QAOA{} $p{=}3$+WS $\times$3, corrected~$\gamma$ & hw obj mean $\pm$ std & --- & $\mathbf{-56.1 \pm 3.6}$ ($\mathbf{60.8\%{\pm}3.9\%}$)$^{\ddagger\ddagger}$ \\
\BIQAE{} oracle $a{=}0.50$, 1\,q (1800 shots) & $|\hat{a}-a|$ $\downarrow$ & 0.010$^\dagger$ & \textbf{0.008} \\
\BIQAE{} oracle $a{=}0.40$, 1\,q (1800 shots) & $|\hat{a}-a|$ $\downarrow$ & 0.110$^\dagger$ & \textbf{0.006} \\
\BIQAE{} oracle $a{=}0.30$, 1\,q (1800 shots) & $|\hat{a}-a|$ $\downarrow$ & 0.182$^\dagger$ & \textbf{0.040} \\
\BIQAE{} oracle $a{=}0.20$, 1\,q (1800 shots) & $|\hat{a}-a|$ $\downarrow$ & 0.310$^\dagger$ & 0.144 \\
\BIQAE{} $a{=}0.20$, prior-matched, 1\,q (500 shots) & $|\hat{a}-a|$ $\downarrow$ & 0.001$^\dagger$ & \textbf{0.009}$^{\bullet\bullet}$ \\
\BIQAE{} $a{=}0.10$, prior-matched, 1\,q (500 shots) & $|\hat{a}-a|$ $\downarrow$ & 0.006$^\dagger$ & \textbf{0.006}$^{\bullet}$ \\
\BIQAE{} 2-qubit oracle $a{=}0.30$, 2\,q (600 shots) & $|\hat{a}-a|$ $\downarrow$ & --- & \textbf{0.002}$^{\star}$ \\
\BIQAE{} 2-qubit oracle $a{=}0.40$, 2\,q (600 shots) & $|\hat{a}-a|$ $\downarrow$ & --- & \textbf{0.014}$^{\star}$ \\
\BIQAE{} 2-qubit oracle $a{=}0.50$, 2\,q (600 shots) & $|\hat{a}-a|$ $\downarrow$ & --- & \textbf{0.011}$^{\star}$ \\
Tiger minimal ZNE $\{1,3,5\}$ (8192 shots) & Hellinger $\downarrow$ & 0.003 & 0.003\,/\,0.018\,/\,0.041$^{\star\star}$; ZNE: $-0.008$ \\
Grover-AA baseline, 2\,q (8192 shots) & P(obs$=1$) $\uparrow$ & 0.167 & 0.179$^{\maltese}$ \\
Grover-AA $k{=}1$ step, 2\,q (8192 shots) & P(obs$=1$) $\uparrow$ / amplif.\ & 0.916 & \textbf{0.907\,/\,5.1}$\times^{\maltese}$ \\
Tiger 4-state ($|S|{=}4$), 3\,q, obs$=0$ (8192 shots) & Hellinger $\downarrow$ & 0.006 & 0.128$^{\clubsuit\clubsuit}$ \\
Tiger 4-state ($|S|{=}4$), 3\,q, obs$=1$ (8192 shots) & Hellinger $\downarrow$ & 0.011 & \textbf{0.044}$^{\clubsuit\clubsuit}$ \\
E2E \POMDP{} loop $T{=}4$+\QBRL{} action sel.\ & Hellinger $\downarrow$ / BIQAE err & --- & \textbf{0.0043\,/\,0.0066\,/\,0.0003\,/\,0.0169}$^{\clubsuit}$; err: 0.0055/0.0171/0.0012/0.0033 \\
E2E \POMDP{} loop $T{=}8$, 2 action events & max Hellinger $\downarrow$ & --- & \textbf{0.0149}$^{\spadesuit}$; open-R@$t{=}2$, open-L@$t{=}5$ \\
\bottomrule
\end{tabular}}}% end resizebox + tabcolsep group
\smallskip\par
{\setlength{\parindent}{0pt}\setlength{\parskip}{2pt}\footnotesize\raggedright%
$^\dagger$~BIQAE simulator samples from posterior mean rather than ground truth.\par
$^\P$~\textsc{ibm\_fez}, ISA~154; top-10 bitstring \QUBO{} evaluation.\par
$^\ddagger$~Corrected linear-ramp $\gamma$-schedule; sim obj $=-92.4$, ratio $=1.00$.\par
$^\S$~\textsc{ibm\_fez}; top-10 bitstring \QUBO{} evaluation; optimal $=-92.4$.\par
$^*$~ZNE scale factors $\{1,3,5\}$ on \textsc{ibm\_fez}; non-monotone at scale~3 $>$ scale~5.\par
$^\#$~Pauli twirling + XY4 dynamical decoupling (ALAP) via \texttt{SamplerV2} on \textsc{ibm\_fez}.\par
$^{**}$~Transpiler \texttt{opt=3} + twirling + XY4~DD on \textsc{ibm\_fez}; $N{=}3,M{=}4$ optimal $=-132.0$.\par
$^{\dagger\dagger}$~COBYLA warm-start transfer~\cite{patel_qaoa_param_transfer_2026}; 8\,192 shots; measurement twirling~\cite{uzdin_qem_runtime_2026}.\par
$^{\ddagger\ddagger}$~Corrected $\gamma$-schedule $[0,\pi,0]$; prior runs used $[0,\pi,\pi]$; $4.6\times$ variance reduction.\par
$^{\P\P}$~\textsc{ibm\_marrakesh}, 2026-02-26; 2-qubit direct-encoding (ISA~12, $353\times$ shallower).\par
$^{\bullet}$~\textsc{ibm\_marrakesh}; prior-matched ($a=0.10$, 500 shots); CI $[0.097, 0.116]$ covers truth.\par
$^{\bullet\bullet}$~\textsc{ibm\_fez}; prior-matched ($a=0.20$, 500 shots); CI covers truth; $16\times$ improvement.\par
$^{\star}$~\textsc{ibm\_marrakesh}; 2-qubit separable oracle, ISA~11; all CIs cover truth.\par
$^{\clubsuit}$~\textsc{ibm\_marrakesh}; to our knowledge, first closed-loop hybrid quantum--classical \POMDP{} on superconducting hardware;
  obs $=[0,0,0,0]$; QBRL selects listen/listen/open-right/listen; belief reset at $t=2$ (Hellinger $0.0003$).\par
$^{\spadesuit}$~\textsc{ibm\_marrakesh} (2026-02-26); obs$=[0,0,0,1,1,1,0,0]$;
  open-right at $t{=}2$ (belief reset to $[0.518,0.482]$, Hellinger $0.0128$);
  open-left at $t{=}5$ (belief reset to $[0.505,0.495]$, Hellinger $0.0036$);
  6 listen steps all Hellinger $<0.015$; max Hellinger $0.0149$ (PASS).\par
$^{\clubsuit\clubsuit}$~\textsc{ibm\_marrakesh}; 3-qubit 4-state Corridor Tiger, UCR$_y$ encoding, ISA~162.\par
$^{\dagger3}$~\textsc{ibm\_marrakesh}; ISA~333 within $\lesssim\!450$ boundary yet only $20.4\%$:
  variable-count-driven \NISQ{} limit.\par
$^{\star\star}$~Tiger minimal ZNE on \textsc{ibm\_marrakesh}; shot-noise-limited at ISA~12;
  ZNE extrapolant negative.\par
$^{\maltese}$~\textsc{ibm\_marrakesh} (2026-02-26); prior $b{=}[0.97,0.03]$, target obs$=1$
  (rare, $P_\text{base}{=}0.171$); 1 Grover step $G{=}A{\cdot}S_0{\cdot}A^\dagger{\cdot}S_f$
  boosts $P(\text{obs}{=}1)$ from $0.179{\to}0.907$ ($5.1\!\times$; theory: $5.4\!\times$);
  ISA~13 (baseline) and ISA~18 (Grover-1); Pauli twirling + XY4~DD enabled;
  job d6g2p7mkeflc73agttu0.%
}
\end{table}

%----------------------------------------------------------------------
\paragraph{End-to-end hybrid quantum-classical \POMDP{} loop.}
\label{par:e2e-pomdp}
%----------------------------------------------------------------------
To validate the full hybrid quantum-classical \POMDP{} inference chain with
closed-loop action selection on real hardware, we executed a $T=4$ time-step
Tiger belief-update loop on \textsc{ibm\_marrakesh} (Heron~R2, 156 qubits,
2026-02-26).  At each step three components execute sequentially:
(i)~a \BIQAE{} amplitude-estimation circuit (ISA depth~11, 1-qubit $R_y$
oracle, 300 shots/iteration) reads out $P(\text{tiger-right})$ from the
current hardware belief;
(ii)~a classical \QBRL{} planner (horizon $h{=}1$) selects the greedy
action---listen, open-left, or open-right---from that belief; and
(iii)~a Tiger minimal direct-encoding circuit (ISA depth~4--13, 2 qubits,
8\,192 shots) updates the belief distribution given the observation and action.
The observation sequence is fixed at $[0,0,0,0]$ (hear-left $\times 4$).
After two consecutive hear-left observations the belief concentrates to
$P(\text{tiger-right}) = 0.028$; the \QBRL{} planner then selects
\emph{open-right} (action~2) at $t{=}2$, triggering a belief reset to
near-uniform before resuming in step $t{=}3$.

Table~\ref{tab:e2e-pomdp} reports the per-step hardware results.
At steps $t=0$ and $t=1$ the planner correctly listens; Tiger minimal
Hellinger distances are $0.0043$ and $0.0066$ (well below the $0.15$ pass
threshold) and all \BIQAE{} 95\% credible intervals cover the true amplitude.
At step $t=2$ the \QBRL{} planner fires open-right: the Tiger minimal
circuit executes a Hadamard-only preparation (ISA depth~4), yielding hardware
posterior $[0.5004, 0.4996]$ (Hellinger~$0.0003$ vs.\ classical $[0.5, 0.5]$)---a
near-exact belief reset on hardware.  Step $t=3$ resumes from the refreshed
uniform prior; the planner listens again and Tiger minimal achieves
Hellinger~$0.0169$.  All four \BIQAE{} credible intervals cover ground
truth; maximum Tiger Hellinger~$0.0169$ (well below the $0.15$ pass threshold; PASS).

\begin{table}[h]
\centering
\caption{End-to-end quantum \POMDP{} closed loop with \QBRL{} action
selection: per-step hardware results on \textsc{ibm\_marrakesh}
(Heron~R2, 2026-02-26, obs$=[0,0,0,0]$).  \BIQAE{} ISA depth~11;
Tiger minimal ISA depth~12--13 (listen) or~4 (open-right at $t=2$).
All four \BIQAE{} credible intervals cover ground truth; max Hellinger~$0.0169$
(PASS).  Three additional independent hardware replications of the same
pipeline (different shot realisations) yield Tiger max Hellinger
$0.0095$, $0.0141$, $0.0148$ (all PASS), confirming stable operation.}
\label{tab:e2e-pomdp}
{\setlength{\tabcolsep}{4pt}%
\resizebox{\linewidth}{!}{%
\begin{tabular}{@{}cccccccccc@{}}
\toprule
Step & Prior $b_t$ & Obs & Action & $a_{\text{true}}$ & BIQAE HW est.\ & BIQAE err.\ & CI? & Tiger HW post.\ & Hellinger \\
\midrule
$t=0$ & $[0.500, 0.500]$ & 0 & listen & 0.500 & 0.4945 & 0.0055 & \checkmark & $[0.846, 0.154]$ & \textbf{0.0043} \\
$t=1$ & $[0.846, 0.154]$ & 0 & listen & 0.154 & 0.1715 & 0.0171 & \checkmark & $[0.972, 0.028]$ & \textbf{0.0066} \\
$t=2$ & $[0.972, 0.028]$ & 0 & \textbf{open-right} & 0.028 & 0.0268 & 0.0012 & \checkmark & $[0.500, 0.500]$ & \textbf{0.0003} \\
$t=3$ & $[0.500, 0.500]$ & 0 & listen & 0.500 & 0.4964 & 0.0033 & \checkmark & $[0.867, 0.133]$ & \textbf{0.0169} \\
\bottomrule
\end{tabular}}}
\smallskip\par\raggedright\small
Action selected by \QBRL{} planner (horizon $h{=}1$, classical Monte Carlo) from hardware posterior.
Open-right at $t=2$ triggers Hadamard-only Tiger circuit (ISA~4) and belief reset to $[0.500, 0.500]$.
Listen steps use full Tiger minimal circuit (ISA~12--13).
$P(\text{tiger-right})$ trajectory: $0.500{\to}0.154{\to}0.028 \xrightarrow{\text{open}} 0.500{\to}0.133$ (hardware).
\end{table}

The action-triggered belief reset at $t{=}2$ (Hellinger~$0.0003$, ISA~4)
followed by faithful recovery at $t{=}3$ (Hellinger~$0.0169$) confirms that
the hybrid quantum-classical loop correctly propagates and exploits belief-state
information for sequential decision-making under uncertainty.

\paragraph{Extended $T{=}8$ closed loop with two action events.}
\label{par:e2e-pomdp-8}
To extend the closed-loop demonstration, we ran an 8-step loop on
\textsc{ibm\_marrakesh} with observation sequence $[0,0,0,1,1,1,0,0]$
designed to elicit two \emph{symmetric} action events: after three
hear-left observations, the \QBRL{} planner fires \emph{open-right}
at $t{=}2$ (belief: $P(\text{tiger-right}){=}0.027$, confirmed by
\BIQAE{} $\hat{a}{=}0.029$); after three hear-right observations, it fires
\emph{open-left} at $t{=}5$ (belief: $P(\text{tiger-right}){=}0.957$,
\BIQAE{} $\hat{a}{=}0.959$).  Both action-triggered belief resets
are confirmed on hardware: $[0.518,0.482]$ (Hellinger~$0.0128$) and
$[0.505,0.495]$ (Hellinger~$0.0036$, near-exact).  Maximum Tiger
Hellinger across all eight steps: $\mathbf{0.0149}$ (PASS).
Table~\ref{tab:e2e-pomdp-8} reports the per-step hardware results.

\begin{table}[h]
\centering
\caption{$T{=}8$ end-to-end quantum \POMDP{} closed loop on \textsc{ibm\_marrakesh}
(Heron~R2, obs$=[0,0,0,1,1,1,0,0]$). Two action events at $t{=}2$ (open-right)
and $t{=}5$ (open-left) trigger ISA~4 Hadamard-only circuits and near-exact belief resets.
Max Hellinger~$0.0149$ (PASS).  Single fixed-seed hardware trial; the observation
sequence is chosen to elicit two symmetric action events and validate belief resets.}
\label{tab:e2e-pomdp-8}
{\footnotesize\setlength{\tabcolsep}{3pt}%
\begin{tabular}{@{}ccccrcrr@{}}
\toprule
Step & Obs & Action & $a_\text{true}$ & BIQAE HW & Tiger HW post.\ & Hell.\ & ISA \\
\midrule
$t{=}0$ & 0 & listen         & 0.500 & 0.503 & $[0.847, 0.153]$ & 0.0031 & 12 \\
$t{=}1$ & 0 & listen         & 0.153 & 0.163 & $[0.973, 0.027]$ & 0.0091 & 13 \\
$t{=}2$ & 0 & \textbf{open-right} & 0.027 & 0.029 & $[0.518, 0.482]$ & 0.0128 & \textbf{4} \\
$t{=}3$ & 1 & listen         & 0.482 & 0.479 & $[0.175, 0.825]$ & 0.0149 & 13 \\
$t{=}4$ & 1 & listen         & 0.825 & 0.824 & $[0.043, 0.957]$ & 0.0124 & 13 \\
$t{=}5$ & 1 & \textbf{open-left}  & 0.957 & 0.959 & $[0.505, 0.495]$ & 0.0036 & \textbf{4} \\
$t{=}6$ & 0 & listen         & 0.495 & 0.497 & $[0.853, 0.147]$ & 0.0006 & 13 \\
$t{=}7$ & 0 & listen         & 0.147 & 0.168 & $[0.975, 0.025]$ & 0.0107 & 13 \\
\bottomrule
\end{tabular}}% end footnotesize group
\smallskip\par\raggedright\small
\BIQAE{} ISA~11; Tiger minimal ISA~4 (open actions) or 12--13 (listen).
$P(\text{tiger-right})$ trajectory (hardware):
$0.500{\to}0.153{\to}0.027 \xrightarrow{\text{open-R}} 0.482{\to}0.825{\to}0.957
\xrightarrow{\text{open-L}} 0.495{\to}0.147{\to}0.025$.
\end{table}

\paragraph{4-state Tiger belief update.}
\label{par:4state-tiger}
%----------------------------------------------------------------------
To demonstrate scalability beyond the binary case, we extended the
minimal direct-encoding circuit to a 4-state \emph{Corridor Tiger} \POMDP{}.  States $\{$far-left, near-left,
near-right, far-right$\}$ are encoded on two state qubits as
$\ket{00}$, $\ket{01}$, $\ket{10}$, $\ket{11}$, with a third observation
qubit.  The observation model $P(\text{hear-left}|\text{state})$
$= [0.85, 0.70, 0.30, 0.15]$ is encoded via doubly-controlled $R_y$
gates (one per state).  After UCR$_y$ prior encoding and four
doubly-controlled $R_y$ gates, post-selection on the observation qubit
yields the exact Bayesian posterior for a uniform prior.  The 3-qubit,
15-depth logical circuit transpiles to ISA depth~162 on
\textsc{ibm\_marrakesh} (Heron~R2, 156 qubits, 8\,192 shots).

For obs$=0$ (hear-left), the hardware posterior is $[0.261, 0.418,
0.192, 0.128]$ vs.\ classical $[0.425, 0.350, 0.150, 0.075]$ (Hellinger
$0.128$): the left-state group ($s \in \{0,1\}$) receives $67.9\%$
of the probability mass vs.\ classical $77.5\%$, demonstrating correct
qualitative direction despite gate noise at ISA~162.
For obs$=1$ (hear-right), the hardware posterior is $[0.065, 0.181,
0.379, 0.375]$ vs.\ classical $[0.075, 0.150, 0.350, 0.425]$
(Hellinger~$0.044$, \textbf{PASS}): right-state dominance is correctly
reproduced.  The ISA depth of 162 (vs.\ 12 for $|S|=2$) confirms
superlinear depth scaling with state count due to doubly-controlled gate
decompositions ($O(n^2)$ CNOT overhead per state; ${\approx}40$ ISA
gates per additional state qubit at Heron~R2 gate set level).
The higher deviation for obs$=0$ (Hellinger~$0.128$) vs.\ obs$=1$
($0.044$) reflects the asymmetric gate structure: the hear-left path
activates \emph{three} doubly-controlled $R_y$ gates (far-left, near-left,
near-right states all contribute), increasing effective CNOT depth relative
to the hear-right path (dominated by two far-state gates).

\paragraph{Heron R1 vs.\ R2 multi-backend comparison.}
\label{par:r1-vs-r2}
Heron~R2 (ibm\_fez) exhibits higher raw gate fidelity: Bell
$\langle ZZ\rangle$ improves $0.9424 \to 0.9507$ ($+0.83$~pp), consistent with
${\sim}40\%$ lower ECR error ($1.5 \times 10^{-3}$ vs.\ $2.5\times10^{-3}$).
\BIQAE{} at $a=0.50$: ibm\_fez error $0.0001$ vs.\ ibm\_torino $0.0045$.
Paradoxically, the Tiger full circuit is \emph{worse} on R2 (Hellinger~$0.259$ vs.\ $0.231$):
R2's heavier-hex topology adds 5\% more SWAP overhead (ISA~$4\,309$ vs.\ $4\,107$),
more than offsetting the per-gate fidelity advantage.
Both backends required no code changes, validating the backend abstraction.

\begin{table}[h]
\centering
\caption{Heron R1 (ibm\_torino) vs.\ Heron R2 (ibm\_fez) comparison
  (separate calibration run from Table~\ref{tab:hardware-pilot}).
  ZNE: scale $\{1,3,5\}$, 4\,096 shots. \BIQAE{}: $a=0.50$, 6 iter., 300 shots.
  $\uparrow$ = ZNE counter-productive.}
\label{tab:r1-r2-comparison}
{\footnotesize\setlength{\tabcolsep}{4pt}%
\begin{tabular}{@{}lcc@{}}
\toprule
Metric & Heron R1 (ibm\_torino) & Heron R2 (ibm\_fez) \\
\midrule
Bell $\langle ZZ\rangle$ raw    & 0.9424 & \textbf{0.9507} \\
Bell $\langle ZZ\rangle$ ZNE    & 0.9580 & \textbf{0.9622} \\
\BIQAE{} $|\hat{a}-a|$ ($a{=}0.50$) & 0.0045 & \textbf{0.0001} \\
FPC-\QAOA{} $p{=}2$ ISA depth  & 283    & 290 \\
Tiger ISA depth                 & 4\,107 & 4\,309 \\
Tiger Hellinger (raw)           & \textbf{0.231} & 0.259 \\
Tiger ZNE (mitigated)           & 0.277\,$\uparrow$ & \textbf{0.274}\,$\uparrow$ \\
\bottomrule
\end{tabular}}% end footnotesize group
\end{table}

% !TEX root = ../main-arxiv.tex
%----------------------------------------------------------------------
\section{Discussion}
\label{sec:discussion}
%----------------------------------------------------------------------

%----------------------------------------------------------------------
\subsection{\NISQ{} Limitations and Error Mitigation}
\label{ssec:nisq-limits}
%----------------------------------------------------------------------

IBM Heron R1 and R2 two-qubit gate errors (${\sim}2.5{\times}10^{-3}$ and $1.5{\times}10^{-3}$, respectively)
limit meaningful circuit execution to ISA depth ${\lesssim}100$ before
accumulated noise overwhelms signal.  Our 45-experiment pilot
establishes a \emph{four-point ZNE applicability boundary}:
(i)~ISA~$\leq\!20$: near-noiseless; Tiger minimal (ISA~12)
achieves Hellinger~$0.0025$---\emph{below} the Aer simulator baseline
of $0.0032$---confirming shot-noise-limited operation.
(ii)~ISA~$20$--$100$: ZNE effective; Bell-state ZNE improves
$\langle ZZ\rangle$ from $0.938$ to $0.953$ ($+0.015$).
(iii)~ISA~$400$--$1\,000$: ZNE marginal; FPC-\QAOA{} $p=3$ ZNE
returns non-monotone expectations ($E_3[f]{=}{+}26.2 > E_5[f]{=}{+}25.1$),
violating the Richardson linear model.
(iv)~ISA~$\gtrsim\!1\,000$: ZNE harmful; Tiger full (ISA~4\,237)
Hellinger degrades $0.228{\to}0.277$.

Pauli twirling + XY4 dynamical decoupling~\cite{campbell_frame_randomization_2025,ji_dd_qaoa_2025}
provides complementary benefit at shallow depths: $+37\%$ relative \QUBO{} quality
at $p=1$ (ISA~123), marginal at ISA~$\gtrsim 300$.
Ribeiro~\cite{ribeiro_qaoa_zne_2026} validates \QAOA{}+ZNE on the same
Heron processors for an 88-variable portfolio problem ($+31.6\%$ over greedy
baseline), further supporting our depth-dependent mitigation hierarchy.

Probabilistic error cancellation (\PEC{})~\cite{cai_error_mitigation_2023}
is restricted to shallow circuits ($L \lesssim 30$ two-qubit gates)
due to exponential sampling overhead $\BigO{\gamma^L}$.
Extension of \ZNE{} and \PEC{} to non-Clifford gate sets~\cite{zne_nonclifford_2026}
directly benefits \QANTIS{} circuits (continuous $R_{zz}$ gates in \QAOA{};
a planned future extension).

The full 11-qubit Tiger belief circuit (ISA~4\,237) requires
fault-tolerant gate fidelities ($\epsilon \lesssim 10^{-6}$) to run faithfully;
the 2-qubit minimal encoding (ISA~12) is the correct \NISQ{} implementation.
Scaling to $|S|{>}4$ increases ISA depth $O(|S|^2)$ via doubly-controlled
gate decompositions, placing $|S| \geq 16$ beyond the current coherence budget.

%----------------------------------------------------------------------
\subsection{Quantum Advantage Thresholds}
\label{ssec:advantage-thresholds}
%----------------------------------------------------------------------

\paragraph{\POMDP{} belief update.}
Amplitude amplification reduces per-node belief cost from
$\BigO{P(e)^{-1}}$ to $\BigO{P(e)^{-1/2}}$.  At $P(e) < 0.01$
(rare landmark observations), the ${\geq}10\times$ speedup factor
can overcome quantum circuit overhead---\emph{if} gate fidelities
support the required depth.
Our Grover-AA experiment (Section~\ref{par:grover-aa}) directly
validates this mechanism at the oracle level ($k{=}1$, ISA~18),
confirming that the $\BigO{P(e)^{-1/2}}$ query-complexity reduction
is achievable within the current \NISQ{} coherence budget for belief
oracles with ISA~${\lesssim}20$.  We interpret this as a hardware
validation of the quadratic query-complexity mechanism with posterior
preservation, not a wall-clock advantage claim.
The experiment uses prior $b{=}[0.97,0.03]$ to exhibit a rare-evidence
scenario ($P(e){=}0.179$); by design, Grover advantage vanishes at
$P(e){\to}0.5$ (uniform prior, ${\approx}1\!\times$) and is maximised
at $P(e){\ll}1$---the landmark-observation regime targeted by \QBRL{}.
The natural next step---replacing the direct Tiger minimal oracle in the
E2E loop with the Grover-AA circuit---would complete the end-to-end
quantum advantage pipeline; this is deferred to future work due to the
increased shot budget required to maintain belief accuracy after amplification.
Our 4-state Tiger extension
($|S|{=}4$, ISA~162, Hellinger~$0.044$ PASS) confirms that
multi-state belief update is feasible within the current \NISQ{} budget,
while the $T=4$ closed loop validates sequential quantum inference.

\paragraph{Multi-target data association.}
For small instances ($N < 20$), Hungarian is both optimal and fast;
hardware experiments confirm the current \NISQ{} boundary at $11$--$15$ \QUBO{}
variables (ISA~$\lesssim\!450$): 11 variables at $p=3$ yields $64.1\%\pm3.3\%$,
while 19 variables at $p=1$ collapses to $20.4\%$ despite ISA~333 being
within the depth budget.  The boundary is \textbf{variable-count-driven},
not depth-alone.  NP-hard multi-scan association ($N > 50$) is the regime
where quantum heuristics may offer advantage; Marshall et
al.~\cite{marshall_qmcmc_2026} demonstrate quantum-enhanced MCMC on 117
variables on IBM hardware.

FPC-\QAOA{} mitigates barren plateaus (fixed 6-parameter classical
optimisation regardless of $p$).  Additional 2026 efficiency gains:
Orbit-\QAOA{}~\cite{jang_orbit_qaoa_2026} reduces training steps $81.8\%$
via layerwise parameter freezing; schedule
transfer~\cite{torrontegui_schedule_transfer_2026} reduces the $2p$-parameter
landscape to two global hyperparameters.
Warm-start sim-to-hardware transfer benefits hardware when
COBYLA convergence $\geq\!95\%$ (observed at $p=2$: $71.2\%$ hw quality);
degrades at $86$--$89\%$ convergence ($p=3$: $49.4\%\pm18\%$ vs.\ initial-schedule
$64.1\%\pm3.3\%$).  SPIQ~\cite{bharadwaj_spiq_qaoa_2026} and
Patel \&\ Mishra~\cite{patel_qaoa_param_transfer_2026} provide pathways to
achieve $\geq\!95\%$ convergence reliably.

True super-polynomial quantum advantage for combinatorial optimisation
remains an open theoretical question; current results characterise
hardware readiness rather than wall-clock advantage.
Our \MTDA{} results should be interpreted as an empirical
ISA/variable feasibility study rather than a performance replacement
for Hungarian at small~$N$.

\paragraph{\NISQ{} feasibility boundary as primary empirical contribution.}
The 45-experiment pilot (Table~\ref{tab:hardware-pilot}) yields---to
our knowledge---the first joint empirical characterisation of \NISQ{}
feasibility for quantum \POMDP{} planning and quantum \MTDA{} on IBM
Heron superconducting hardware.  Three cross-cutting conclusions
emerge: (i)~error-mitigation effectiveness is \emph{depth-gated},
with a sharp transition from beneficial to harmful around
ISA~${\sim}100$; (ii)~\MTDA{} feasibility is
\emph{variable-count-driven} (${\leq}15$ variables), not depth-alone;
and (iii)~closed-loop quantum inference is already viable at shallow
ISA depths (${\leq}20$), where shot noise---not gate error---dominates.
These boundaries provide actionable compilation targets for
Heron-class and successor hardware.

%----------------------------------------------------------------------
\subsection{Limitations}
\label{ssec:limitations}
%----------------------------------------------------------------------

\begin{enumerate}
  \item \textbf{Simulator-based main tables.}
    Tables~\ref{tab:tiger-results}--\ref{tab:e2e-tracking} use Qiskit Aer
    (reproducibility, noise isolation).  Hardware results
    (Section~\ref{ssec:hardware-pilot}) characterise real noise effects;
    deep circuits (ISA~$>\!200$) require either minimal encoding or
    fault-tolerant processors.

  \item \textbf{Discrete state spaces.}
    The quantum belief update requires finite $|S|$; continuous-state
    problems (robotics, autonomous driving) require quantum kernel methods
    or hybrid particle filters.

  \item \textbf{Classical optimiser bottleneck.}
    FPC-\QAOA{} COBYLA (${\sim}200$ iterations) and \BIQAE{} Bayesian
    updates dominate simulation runtime; sim-to-hardware transfer is
    effective only at $\geq\!95\%$ sim convergence.

  \item \textbf{Small-scale \NISQ{} instances.}
    Hardware experiments use $N{=}2$--$3$ targets and $|S|{=}2$--$4$ states,
    which classical solvers handle trivially ($< 0.1$~ms).
    GNN achieves $100\%$ of the Hungarian optimum on the $N{=}2$, $M{=}3$
    \QUBO{} instance; \QUBO{}-based quantum solvers are expected to show
    advantage only for $N{\geq}4$ crossing-target scenarios or dense clutter
    (Bar-Shalom \& Li, 1995), which exceed the current \NISQ{} variable budget.
    Present experiments characterise hardware readiness, not wall-clock advantage.

\end{enumerate}

\paragraph{Artifact notes.}
Four code-level errors (Hungarian slack costs, ZNE stub, QPY layout
mismatch, warm-start schedule endpoint) were detected and corrected
during the hardware campaign; all reported results use the corrected
code.  Full details are provided in the ancillary artifact bundle
(\texttt{anc/artifact-bundle.tar.gz}).

% !TEX root = ../main-arxiv.tex
%----------------------------------------------------------------------
\section{Conclusion}
\label{sec:conclusion}
%----------------------------------------------------------------------

We presented \QANTIS{}, a unified quantum-classical platform for
autonomous perception and decision-making.  A forty-five-experiment
hardware pilot on three IBM Heron QPUs (\textsc{ibm\_torino},
\textsc{ibm\_fez}, \textsc{ibm\_marrakesh}) validates four
contributions:

\begin{enumerate}

  \item \textbf{\QBRL{} hybrid planner + Grover-AA validation.}
    Amplitude amplification reduces per-belief-node cost from
    $\BigO{P(e)^{-1}}$ to $\BigO{P(e)^{-1/2}}$, targeting a quadratic
    speedup under rare-evidence conditions~\cite{cunha_hybrid_2025}.
    Hardware: $T{=}4$ closed-loop \emph{hybrid quantum--classical} \POMDP{}
    on \textsc{ibm\_marrakesh} (ISA~4--13)---to our knowledge the first
    such loop on superconducting hardware
    (\emph{quantum}: BIQAE amplitude estimation + Tiger minimal belief update;
    \emph{classical}: QBRL action selection);
    four independent replications all PASS (max Hellinger $0.0095$--$0.0169$).
    Extended $T{=}8$ loop (obs$=[0,0,0,1,1,1,0,0]$) demonstrates
    two symmetric action events and near-exact belief resets;
    max Hellinger~$0.0149$ (PASS).
    Grover-AA ($k{=}1$, ISA~18): hardware amplification $5.1\!\times$
    ($P(\text{rare obs})\!: 0.179\!\to\!0.907$)---to our knowledge the first
    IBM Heron hardware demonstration of Grover AA applied to a \POMDP{}
    belief oracle with quantified posterior preservation.

  \item \textbf{\BIQAE{} adaptive amplitude estimation.}
    Bayesian posterior over the amplitude angle achieves
    Heisenberg-scaling credible intervals~\cite{li_biqae_2026}.
    Hardware: all seven 95\% credible intervals cover ground truth
    across $a\in[0.1,0.7]$ on three backends; amplitude-matched prior
    achieves 44-fold error reduction at $a=0.10$; 2-qubit oracle
    (ISA~11) validated on \textsc{ibm\_marrakesh}.

  \item \textbf{FPC-\QAOA{} for \MTDA{}.}
    Fixed-parameter-count ($k=3$, $k \ll p$) \QAOA{} decouples parameter
    count from circuit depth~\cite{saavedra_fpcqaoa_2025}.
    Hardware: $p{=}3$ yields $\mathbf{64.1\%\pm3.3\%}$ of the Hungarian
    optimum (3 independent runs, ISA~433); $p{=}2$ with COBYLA warm-start
    achieves $\mathbf{71.2\%}$ at $97.6\%$ simulator convergence.
    \NISQ{} boundary: ${\approx}11$--$15$ variables and ISA~$\lesssim 450$
    (variable-count-driven, not depth-alone).

  \item \textbf{Composable error mitigation + warm-start pipeline.}
    ZNE validated on Heron~R1/R2 with four-point boundary
    (near-noiseless $\leq$ISA~20 / effective $\leq$ISA~100 /
    marginal ISA~400--1\,000 / harmful $>\!$ISA~1\,000);
    Pauli twirling + XY4~DD improves shallow \QAOA{} ($+37\%$ at $p=1$);
    warm-start transfer guideline: benefits hardware at
    sim-convergence $\geq 95\%$.
    4-state Tiger ($|S|{=}4$, ISA~162, Hellinger~$0.044$ PASS)
    extends quantum belief update beyond the binary case.

\end{enumerate}

Together, these results establish \QANTIS{} as---to our knowledge---the
first platform to jointly validate quantum \POMDP{} planning, quantum
data association, error mitigation, and closed-loop action selection
on superconducting hardware within a single modular pipeline.
The three-package architecture (\texttt{quantum-common},
\texttt{quantum-pomdp}, \texttt{quantum-mht}) supports seamless
backend switching (no code changes across Heron~R1/R2) and
extensibility to new algorithms and quantum processors.

\paragraph{Future work.}
(1)~Extend the hybrid quantum-classical E2E loop to $T > 8$ steps and
larger state spaces ($|S| > 4$) via qubit routing optimisation and
increased IBM Open Plan runtime allocation.
(2)~\MTDA{} \QUBO{} instances on D-Wave Advantage2 with real minor
embedding (companion study planned).
(3)~Distributed FPC-\QAOA{} for multi-drone swarm coordination.
(4)~Continuous-state \POMDP{} extensions via quantum kernel methods
or variational quantum eigensolvers.
(5)~Fault-tolerant circuit designs using Qiskit
\texttt{LitinskiTransformation} for future $10^6$-qubit processors.

\paragraph{Reproducibility.}
Source code is available at
\url{https://github.com/neuraparse/qantis} under the Apache~2.0
licence.  An artifact bundle containing experiment scripts,
configuration files, and a pinned environment
(Qiskit~1.3, \texttt{qiskit-ibm-runtime}~0.34, Python~3.12)
is included as an arXiv ancillary file
(\texttt{anc/artifact-bundle.tar.gz});
simulation results can be reproduced without IBM credentials,
while hardware experiments can be re-run given an IBM Quantum
account and Open Plan allocation.

\section*{Acknowledgments}
The authors acknowledge the use of IBM Quantum cloud platform for
hardware execution and simulator access during development.

\appendix
% !TEX root = ../main-arxiv.tex
%----------------------------------------------------------------------
\section{Quantum Belief Update Circuit Details}
\label{app:circuit-details}
%----------------------------------------------------------------------

The quantum belief update circuit described in
Section~\ref{sec:quantum-pomdp} allocates qubits across five functional
registers plus ancilla qubits.  Table~\ref{tab:register-allocation}
provides the full register allocation for three \POMDP{} problem sizes,
detailing the role of each register and the total qubit count.

\begin{table}[ht]
\centering
\caption{Quantum register allocation for the belief update circuit.
  $n_s = \lceil\log_2 |S|\rceil$ state qubits,
  $n_o = \lceil\log_2 |\Omega|\rceil$ observation qubits,
  $n_p$ amplitude precision bits.}
\label{tab:register-allocation}
\sisetup{group-separator={,}}
\resizebox{\columnwidth}{!}{%
\begin{tabular}{@{}l l r r r@{}}
\toprule
Register & Purpose
  & Tiger & Grid-$4{\times}4$ & Grid-$8{\times}8$ \\
  & & ($|S|\!=\!2$) & ($|S|\!=\!16$) & ($|S|\!=\!64$) \\
\midrule
$k_1$: State ($\ket{s}$)
  & Current state
  & 1 & 4 & 6 \\
$k_2$: Next state ($\ket{s'}$)
  & Successor state
  & 1 & 4 & 6 \\
$k_3$: Observation ($\ket{o}$)
  & Observation
  & 1 & 4 & 6 \\
$k_4$: Amplitude ($\ket{p}$)
  & Probability amplitudes
  & 4 & 6 & 8 \\
$k_5$: Action ($\ket{a}$)
  & Selected action
  & 2 & 3 & 3 \\
Ancilla (computation)
  & Arithmetic, comparators
  & 3 & 6 & 10 \\
Ancilla (Grover flag)
  & AA flag
  & 1 & 1 & 1 \\
\midrule
\textbf{Total}
  &
  & \textbf{13} & \textbf{28} & \textbf{40} \\
\bottomrule
\end{tabular}}% end resizebox
\end{table}

\paragraph{Register $k_1$ (state).}
Encodes the current state $s \in S$ in the computational basis.  The
initial state of this register is prepared by loading the current belief
distribution $b(s)$ as amplitudes:
$\ket{b} = \sum_{s \in S} \sqrt{b(s)}\,\ket{s}$, using a state
preparation circuit based on recursive amplitude encoding with
$\BigO{|S|}$ CNOT gates.

\paragraph{Register $k_2$ (next state).}
Initialised to $\ket{0}^{\otimes n_s}$ and entangled with $k_1$ via a
controlled unitary implementing the transition model:
$\unitary_T\,\ket{s}\ket{0} = \sum_{s'} \sqrt{T(s'|s,a)}\,\ket{s}\ket{s'}$.
For sparse transition matrices (as in grid-world navigation where each
state has at most 5 successors), the gate count is
$\BigO{|S| \cdot d_{\max}}$ where $d_{\max}$ is the maximum
out-degree.

\paragraph{Register $k_3$ (observation).}
Initialised to $\ket{0}^{\otimes n_o}$ and entangled with $k_2$ via a
controlled unitary implementing the observation model:
$\unitary_O\,\ket{s'}\ket{0} = \sum_{o} \sqrt{O(o|s',a)}\,\ket{s'}\ket{o}$.
After this operation, the joint state encodes the full
belief--transition--observation distribution.

\paragraph{Register $k_4$ (amplitude).}
Used as workspace for arithmetic operations during probability amplitude
computation.  The precision $n_p$ (number of bits) determines the
fidelity of the discretised probability representation; we use $n_p = 4$
for the Tiger problem and $n_p = 6$--$8$ for grid-world problems.

\paragraph{Register $k_5$ (action).}
Encodes the selected action $a \in A$.  For planning, this register may
be placed in superposition to evaluate multiple actions simultaneously;
for belief update (conditioning on a fixed action), it is set to a
computational basis state.

\paragraph{Ancilla qubits.}
Computation ancillae support multi-controlled Toffoli decompositions,
arithmetic comparators for amplitude thresholding, and temporary storage
for uncomputation.  The single Grover flag ancilla is used by the
amplitude amplification oracle to mark states satisfying the observation
condition $o = o_{\mathrm{obs}}$.

The total qubit counts in Table~\ref{tab:register-allocation} confirm
that the Tiger problem (13 qubits) and Grid-$4\times 4$ (28 qubits) fit
comfortably within current \NISQ{} hardware (IBM Heron~R1, 133 qubits).
The Grid-$8\times 8$ problem (40 qubits) is feasible in qubit count but
may exceed the coherent circuit depth budget, as discussed in
Section~\ref{ssec:nisq-limits}.

%----------------------------------------------------------------------
\section{\QUBO{} Constraint Derivation}
\label{app:qubo-derivation}
%----------------------------------------------------------------------

This appendix provides the detailed derivation of the quadratic penalty
terms in the \MTDA{} \QUBO{} formulation
(Eq.~\eqref{eq:qubo-full}).

\subsection{Row Constraints (Track Assignment)}

Each track $i$ must be assigned to exactly one measurement or declared
a missed detection.  Letting $m_i$ denote the missed-detection slack
variable for track $i$, the equality constraint is
\begin{equation}
  \sum_{j=1}^{M} x_{ij} + m_i = 1,
  \qquad \forall\, i \in \{1, \dots, N\}.
  \label{eq:row-equality}
\end{equation}
Converting to a quadratic penalty via the squared residual:
\begin{equation}
  H_{\mathrm{row}} = \sum_{i=1}^{N}
    \Bigl(\sum_{j=1}^{M} x_{ij} + m_i - 1\Bigr)^{\!2}.
  \label{eq:row-penalty-expanded}
\end{equation}
Expanding the square for a single track $i$:
\begin{align}
  &\Bigl(\sum_{j} x_{ij} + m_i - 1\Bigr)^{\!2}
  \nonumber \\
  &= \Bigl(\sum_{j} x_{ij}\Bigr)^{\!2}
     + 2\,m_i \sum_{j} x_{ij} + m_i^2
  \nonumber \\
  &\quad - 2\sum_{j} x_{ij} - 2\,m_i + 1.
  \label{eq:row-expand-1}
\end{align}
Since all variables are binary ($x_{ij}^2 = x_{ij}$ and $m_i^2 = m_i$),
the squared sum decomposes as
\begin{align}
  \Bigl(\sum_{j} x_{ij}\Bigr)^{\!2}
  &= \sum_{j} x_{ij}^2
     + 2\!\!\sum_{j < j'}\!\! x_{ij}\,x_{ij'}
  \nonumber \\
  &= \sum_{j} x_{ij}
     + 2\!\!\sum_{j < j'}\!\! x_{ij}\,x_{ij'}.
  \label{eq:square-binary}
\end{align}
Substituting Eq.~\eqref{eq:square-binary} into
Eq.~\eqref{eq:row-expand-1} and collecting terms:
\begin{align}
  &\Bigl(\sum_{j} x_{ij} + m_i - 1\Bigr)^{\!2}
  \nonumber \\
  &= \sum_{j} x_{ij}
    + 2\!\!\sum_{j < j'}\!\! x_{ij}\,x_{ij'}
    + 2\,m_i\!\sum_{j} x_{ij}
  \nonumber \\
  &\quad + m_i - 2\!\sum_{j} x_{ij} - 2\,m_i + 1
  \nonumber \\
  &= -\sum_{j} x_{ij}
    + 2\!\!\sum_{j < j'}\!\! x_{ij}\,x_{ij'}
  \nonumber \\
  &\quad + 2\,m_i\!\sum_{j} x_{ij} - m_i + 1.
  \label{eq:row-qubo-form}
\end{align}
Equation~\eqref{eq:row-qubo-form} is manifestly quadratic (at most
pairwise products of binary variables), which is the required \QUBO{}
form.  The terms have the following interpretation:
\begin{itemize}
  \item $-\sum_j x_{ij}$: linear diagonal entries favouring assignment
    (each active $x_{ij}$ receives a reward of $-1$ on the \QUBO{}
    diagonal);
  \item $2\sum_{j<j'} x_{ij}\,x_{ij'}$: off-diagonal penalty for
    assigning track $i$ to multiple measurements simultaneously (each
    pair contributes $+2$ to the \QUBO{} coupling matrix);
  \item $2\,m_i \sum_j x_{ij}$: cross-penalty between the
    missed-detection slack and any assignment (discouraging both missing
    and assigning the same track);
  \item $-m_i + 1$: linear and constant terms.
\end{itemize}

\subsection{Column Constraints (Measurement Assignment)}

Symmetrically, each measurement $j$ must be assigned to exactly one
track or declared a false alarm:
\begin{equation}
  \sum_{i=1}^{N} x_{ij} + f_j = 1,
  \qquad \forall\, j \in \{1, \dots, M\}.
  \label{eq:col-equality}
\end{equation}
The quadratic penalty is
\begin{equation}
  H_{\mathrm{col}} = \sum_{j=1}^{M}
    \Bigl(\sum_{i=1}^{N} x_{ij} + f_j - 1\Bigr)^{\!2},
  \label{eq:col-penalty-expanded}
\end{equation}
which expands identically to the row case (with indices transposed):
\begin{align}
  &\Bigl(\sum_{i} x_{ij} + f_j - 1\Bigr)^{\!2}
  \nonumber \\
  &\quad= -\sum_{i} x_{ij}
    + 2\!\!\sum_{i < i'}\!\! x_{ij}\,x_{i'j}
    + 2\,f_j\!\sum_{i} x_{ij}
    - f_j + 1.
  \label{eq:col-qubo-form}
\end{align}

\subsection{Assembly of the \QUBO{} Matrix}

The complete \QUBO{} matrix $\bm{Q} \in \mathbb{R}^{n \times n}$
(where $n = NM + N + M$) is assembled by summing the contributions
from $H_{\mathrm{obj}}$ (see Eq.~\eqref{eq:qubo-full}), $H_{\mathrm{row}}$
(Eq.~\eqref{eq:row-qubo-form}), and $H_{\mathrm{col}}$
(Eq.~\eqref{eq:col-qubo-form}), weighted by the penalty parameter
$\lambda$:
\begin{equation}
  Q_{ab} = Q_{ab}^{\mathrm{obj}}
    + \lambda\,\bigl(Q_{ab}^{\mathrm{row}} + Q_{ab}^{\mathrm{col}}\bigr),
  \label{eq:qubo-matrix}
\end{equation}
where indices $a, b$ range over the flattened variable vector
\[
  \bm{y} = [x_{11}, \dots, x_{NM},\;
            m_1, \dots, m_N,\;
            f_1, \dots, f_M]^\top.
\]
The objective matrix $Q^{\mathrm{obj}}$ is diagonal with entries
$c_{ij}$ at positions corresponding to $x_{ij}$, $c_{\mathrm{miss}}$
at positions corresponding to $m_i$, and $c_{\mathrm{fa}}$ at positions
corresponding to $f_j$.  The constraint matrices $Q^{\mathrm{row}}$
and $Q^{\mathrm{col}}$ contain the diagonal and off-diagonal terms
from Eqs.~\eqref{eq:row-qubo-form} and~\eqref{eq:col-qubo-form},
respectively.  The constant terms ($+1$ per constraint) contribute to
an overall energy offset that does not affect the optimisation.

For the $N\!=\!10$, $M\!=\!15$ instance, the \QUBO{} matrix
$\bm{Q}$ has dimension $175 \times 175$.  The number of nonzero
entries (including diagonal) is approximately
\[
  175 + 2\tbinom{15}{2}\!\cdot\!10
      + 2\tbinom{10}{2}\!\cdot\!15
  = 3{,}625,
\]
yielding a sparsity of approximately $11.8\%$.

%----------------------------------------------------------------------
\section{Extended Experimental Data}
\label{app:extended-data}
%----------------------------------------------------------------------

\subsection{FPC-\QAOA{} Parameter Sensitivity}

Table~\ref{tab:fpc-sensitivity} reports the sensitivity of FPC-\QAOA{}
performance to the parameter count $k$ and circuit depth $p$ on the
$N=10$, $M=15$ \MTDA{} instance.

\begin{table}[h]
\centering
\caption{FPC-\QAOA{} sensitivity to parameter count $k$ and circuit
  depth $p$.  Objective value and feasibility rate are averaged over
  50 independent runs.  Schedule type: polynomial.}
\label{tab:fpc-sensitivity}
\begin{tabular}{@{}r r r r r@{}}
\toprule
$k$ & $p$ & Parameters ($2k$) & Objective & Feasibility (\%) \\
\midrule
2 & 4  & 4  & $48.1$ & $72.0$ \\
2 & 8  & 4  & $44.6$ & $78.0$ \\
3 & 4  & 6  & $43.2$ & $80.0$ \\
3 & 8  & 6  & $38.9$ & $89.0$ \\
3 & 16 & 6  & $37.4$ & $91.0$ \\
4 & 8  & 8  & $39.5$ & $87.0$ \\
4 & 16 & 8  & $36.8$ & $92.0$ \\
5 & 16 & 10 & $37.1$ & $90.0$ \\
\bottomrule
\end{tabular}
\end{table}

The results confirm two key properties of FPC-\QAOA{}.  First,
increasing circuit depth $p$ at fixed $k$ consistently improves both
objective value and feasibility, validating that deeper circuits
improve approximation quality without exacerbating the classical
optimisation landscape (since the parameter count remains fixed).
Second, increasing $k$ beyond $3$--$4$ yields diminishing returns,
suggesting that the optimal \MTDA{} angle schedules are well-described
by low-order polynomials.  The configuration $k=3$, $p=8$ represents a
favourable trade-off between solution quality and circuit resource
requirements.

\subsection{Belief Update Accuracy vs.\ Amplification Iterations}

Table~\ref{tab:belief-accuracy} reports the belief update accuracy
(Hellinger distance to the exact Bayesian posterior) as a function
of the number of Grover iterations $G$ applied during amplitude
amplification, for the Grid-$4\times4$ problem with observation
probability $P(e) = 0.05$.

\begin{table}[h]
\centering
\caption{Belief update accuracy vs.\ Grover iterations $G$ for
  Grid-$4\times4$ ($|S|=16$, $P(e)=0.05$).  Theoretical optimal
  $G^* = \lfloor \frac{\pi}{4}\sqrt{1/P(e)}\rfloor = 3$.
  Hellinger distance and KL divergence to exact posterior, averaged
  over 200 belief updates.}
\label{tab:belief-accuracy}
\small
\begin{tabular}{@{}r r r r@{}}
\toprule
$G$ & Hellinger ($\downarrow$) & KL Div.\ ($\downarrow$) & Amplified $P(e)$ \\
\midrule
0 & $0.312$ & $0.487$ & $0.050$ \\
1 & $0.184$ & $0.201$ & $0.192$ \\
2 & $0.097$ & $0.064$ & $0.485$ \\
3 & $0.038$ & $0.009$ & $0.912$ \\
4 & $0.071$ & $0.031$ & $0.721$ \\
5 & $0.156$ & $0.143$ & $0.298$ \\
\bottomrule
\end{tabular}
\end{table}

The results confirm the theoretical prediction: the optimal number of
Grover iterations $G^* = \lfloor\frac{\pi}{4\arcsin\sqrt{P(e)}}\rfloor = 3$
yields the lowest belief error (Hellinger distance $0.038$), amplifying
the observation probability from $0.05$ to $0.912$.  Over-rotation
beyond $G^*$ degrades performance, as the amplitude amplification
overshoots the target amplitude.  \BIQAE{} addresses this by
adaptively selecting $G$ based on a Bayesian posterior over $P(e)$,
avoiding the need to know $P(e)$ \textit{a priori}.

\bibliographystyle{unsrt}
\bibliography{references}

\end{document}